\documentclass[twocolumn, preprint]{aastex63}
\usepackage{amsmath,amstext}
\usepackage[T1]{fontenc}
\usepackage{apjfonts} 
\usepackage{natbib}
\usepackage{microtype}
\usepackage{longtable}
\usepackage{verbatim}
\usepackage{upgreek}
\usepackage{hyperref}
\received{July 5, 2021}
\revised{September 3, 2021}
\accepted{September 13, 2021}
\submitjournal{ApJ}

\shorttitle{Rapid outside-in environmental quenching in $z\sim1$ clusters}
\shortauthors{Matharu et al.}
\graphicspath{{./}{figures/}}

\begin{document}

\title{HST/WFC3 grism observations of $z\sim1$ clusters: Evidence for rapid outside-in environmental quenching from spatially resolved H$\upalpha$ maps}

\correspondingauthor{Jasleen Matharu}
\email{jmatharu@tamu.edu}

\author[0000-0002-7547-3385]{Jasleen Matharu}
\affiliation{Department of Physics and Astronomy, Texas A\&M University, College Station, TX, 77843-4242, USA\\}
\affiliation{George P.\ and Cynthia Woods Mitchell Institute for
 Fundamental Physics and Astronomy, Texas A\&M University, College
 Station, TX, 77845-4242, USA\\}
\author[0000-0002-9330-9108]{Adam Muzzin}
\affiliation{Department of Physics and Astronomy, York University, 4700 Keele Street, Toronto, ON, M3J 1P3, Canada\\}
\author[0000-0003-2680-005X]{Gabriel B. Brammer}
\affiliation{Cosmic Dawn Center, Niels Bohr Institute, University of Copenhagen, Jagtvej 128, 2200 Copenhagen N, Denmark\\}
\author[0000-0002-7524-374X]{Erica J. Nelson}
\affiliation{Department for Astrophysical and Planetary Science, University of Colorado, Boulder, CO 80309, USA\\}
\author{Matthew W. Auger}
\affiliation{Institute of Astronomy, University of Cambridge, Madingley Road, Cambridge, CB3 0HA, UK\\}
\author[0000-0002-6528-1937]{Paul C. Hewett}
\affiliation{Institute of Astronomy, University of Cambridge, Madingley Road, Cambridge, CB3 0HA, UK\\}
\author[0000-0003-1535-2327]{Remco van der Burg}
\affiliation{European Southern Observatory, 85748, Garching bei M{\"u}nchen, Germany\\}
\author[0000-0003-4849-9536]{Michael Balogh}
\affiliation{Department of Physics and Astronomy, University of Waterloo, Waterloo, ON, N2L 3G1, Canada\\}
\affiliation{Waterloo Centre for Astrophysics, University of Waterloo, Waterloo, ON, N2L 3G1, Canada\\}
\author[0000-0003-3921-2177]{Ricardo Demarco}
\affiliation{Departamento de Astronom\'ia, Facultad de Ciencias F\'isicas y Matem\'aticas,
Universidad de Concepci\'on, Concepci\'on, Chile\\}
\author[0000-0001-9002-3502]{Danilo Marchesini}
\affiliation{Department of Physics \& Astronomy, Tufts University, 574 Boston Avenue Suites 304, Medford, MA 02155, USA\\}
\author[0000-0003-1832-4137]{Allison G. Noble}
\affiliation{School of Earth and Space Exploration, Arizona State University, Tempe, AZ 85287-1404, USA\\}
\author[0000-0001-5851-1856]{Gregory Rudnick}
\affiliation{Department of Physics and Astronomy, The University of Kansas, Malott room 1082, 1251 Wescoe Hall Drive, Lawrence, KS 66045, USA\\}
\author[0000-0002-5027-0135]{Arjen van der Wel}
\affiliation{Sterrenkundig Observatorium, Department of Physic and Astronomy, Universiteit Gent, Krijgslaan 281 S9, B-9000 Gent, Belgium\\}
\author[0000-0002-6572-7089]{Gillian Wilson}
\affiliation{Department of Physics and Astronomy, University of California Riverside, 900 University Avenue, Riverside, CA 92521, USA\\}
\author{Howard K.C. Yee}
\affiliation{Department of Astronomy and Astrophysics, University of Toronto, 50 St. George Street, Toronto, ON, M5S 3H4, Canada}




\begin{abstract}

We present and publicly release (\url{https://www.gclasshst.com}) the first spatially resolved H$\upalpha$ maps of star-forming cluster galaxies at $z\sim1$, made possible with the Wide Field Camera 3 (WFC3) G141 grism on the Hubble Space Telescope (HST). Using a similar but updated method to 3D-HST in the field environment, we stack the H$\upalpha$ maps in bins of stellar mass, measure the half-light radius of the H$\upalpha$ distribution and compare it to the stellar continuum. The ratio of the H$\upalpha$ to stellar continuum half-light radius, $R[\mathrm{H}\upalpha/\mathrm{C}]=\frac{R_{\mathrm{eff, H}\upalpha}}{R_{\mathrm{eff, Cont}}}$, is smaller in the clusters by $(6\pm9)\%$, but statistically consistent within $1\sigma$ uncertainties. A negligible difference in $R[\mathrm{H}\upalpha/\mathrm{C}]$ with environment is surprising, given the higher quenched fractions in the clusters relative to the field. We postulate that the combination of high quenched fractions and no change in $R[\mathrm{H}\upalpha/\mathrm{C}]$ with environment can be reconciled if environmental quenching proceeds rapidly. We investigate this hypothesis by performing similar analysis on the spectroscopically-confirmed recently quenched cluster galaxies. 87\% have H$\upalpha$ detections, with star formation rates $8\pm1$ times lower than star-forming cluster galaxies of similar stellar mass. Importantly, these galaxies have a $R[\mathrm{H}\upalpha/\mathrm{C}]$ that is $(81\pm8)\%$ smaller than coeval star-forming field galaxies at fixed stellar mass. This suggests the environmental quenching process occurred outside-in. We conclude that disk truncation due to ram-pressure stripping is occurring in cluster galaxies at $z\sim1$, but more rapidly and/or efficiently than in $z\lesssim0.5$ clusters, such that the effects on $R[\mathrm{H}\upalpha/\mathrm{C}]$ become observable just after the cluster galaxy has recently quenched.

\end{abstract}



\section{Introduction} 
\label{sec:intro}
It has been known for a long time that cluster galaxies are on average redder in colour, less actively star-forming and more bulge-dominated in morphology compared to galaxies in the low-density field environment \citep{Abell1965,OemlerAugustus1974,Dressler1980,Postman1984,Balogh1997,Poggianti1999,Lewis2002,Gomez2003a,Postman2005}.

In more recent times, it has been found that even at fixed stellar mass, the fraction of galaxies with quenched star formation is higher in high-density regions such as clusters, both at low ($z\sim0$) and high ($z\sim1$) redshift \citep{Kauffmann2004,Balogh2004b,Poggianti2006,Cooper2006,Cooper2007,Kimm2009,VonderLinden2010,Peng2010d,Muzzin2012,Mok2013,Davies2016,Kawinwanichakij2016a,Nantais2016,Nantais2017,Guglielmo2019,Pintos-Castro2019,VanderBurg2020}. Population studies such as these provide us with the ability to constrain global properties of galaxies residing in different environments, such as star formation rates, quenched fractions and quenching timescales. However, the statistical power of these studies do not allow for strong constraints on the physics that drives these global trends.

In the case of star formation and its quenching, the strongest probe allowing us to understand how it operates in galaxies is spatial information. Early results using spatial information were obtained using narrow-band imaging, where different wavelengths tracing star formation operating on different timescales were exploited. A common tracer used for instantaneous ($\sim10$~Myr) star formation is H$\upalpha$ emission. The ultraviolet radiation emitted by young, massive O and B stars leads to recombination and therefore emission in H$\upalpha$ \citep{Kennicutt1998}. When the spatial distribution of H$\upalpha$ emission is compared to the spatial distribution of the integrated star formation history for a galaxy (the stellar continuum), it allows us to compare {\it where} star formation has been occurring in the past $\sim10$~Myr to where star formation was occurring in the more distant past. Using this technique in the local Universe, \cite{Koopmann2004a} found that more than half of spiral galaxies in the Virgo cluster have truncated H$\upalpha$ discs relative to their {\it R}-band discs, compared to only $\sim10\%$ in the low density field environment (See also \citealt{Hodge1983,Athanassoula1993,Ryder1994,Kenney1999,Koopmann2004,Koopmann2006,Cortes2006,Crowl2006,Abramson2011a,Vollmer2012,Gavazzi2013,Kenney2015,Abramson2016b,Lee2017,Gavazzi2018,Cramer2019,Boselli2020}). Studies such as these allowed us to explain how environmental quenching operates at $z\sim0$: it seems to destroy disks from the ``outside-in", making galaxies appear smaller with a more concentrated light profile.

Recently, with the emergence of integral field spectroscopy, studies on environmental quenching have become more sophisticated. We now have the ability to make statements on how the kinematics of galaxies are affected by environmental quenching processes such as ram-pressure stripping (e.g. GASP, \citealt{Poggianti2017,Bellhouse2017,Gullieuszik2017,Fritz2017,Moretti2018a,Jaffe2018,Vulcani2018, Poggianti2019, Poggianti2019a, Vulcani2020} and others, \citealt{Fumagalli2014a,Fossati2016a,Consolandi2017}). Similarly, new deeper narrow-band imaging surveys combined with long-slit spectroscopy have made it possible to measure quenching timescales as a function of galactocentric radius (e.g. VESTIGE, \citealt{Fossati2018,Boselli2020,Boselli2021}). For the first time, we are able to quantify the rate at which specific environmental quenching mechanisms operate and where they start and finish acting in a galaxy.

However, the aforementioned progress in our understanding of environmental quenching is confined to the low redshift Universe. Naturally, such detailed observations are challenging to obtain at high redshift, but arguably more necessary. To make statements on which environmental quenching mechanism dominates the production of the many red, dead and bulge-dominated galaxies in today's galaxy clusters, we must look to high redshifts, where galaxy clusters were still in the process of forming, cosmic star formation rates were higher (\citealt{Madau2014} and references therein) and galaxies can be observed directly whilst in the process of quenching.

In a first study of its kind, \cite{Nelson2012} used {\it Hubble Space Telescope} (HST) {\it Wide Field Camera ${\it 3}$} (WFC$3$) slitless spectroscopy to construct spatially resolved H$\upalpha$ maps of star-forming field galaxies at $z\sim1$ as part of the 3D-HST survey. By comparing the spatial extent of the H$\upalpha$ emission to the stellar continuum, \cite{Nelson2015} were able to conclude that star-forming field galaxies are growing in an ``inside-out" fashion via star formation, since their H$\upalpha$ emission is more spatially extended than their stellar continuum at all stellar masses. More recently as part of the KMOS-CLASH survey, \cite{Vaughan2020} demonstrated the power of spatially resolved information from integral field spectroscopy, finding that cluster galaxies at $z\sim0.5$ have $(26\pm12)\%$ smaller H$\upalpha$ to stellar continuum size ratios than coeval field galaxies. This result provided direct evidence of outside-in environmental quenching outside the local Universe. In this paper, we use the same technique as \cite{Nelson2015} on the largest sample of spectroscopically-confirmed cluster galaxies at $z\sim1$ in an attempt to make the first direct measurement of environmental quenching at high redshift.

We do this by measuring the spatial extent of their H$\upalpha$ emission and comparing it to the spatial extent of their stellar component, using the low redshift technique discussed earlier that has been shown to be a powerful tool in constraining environmental quenching mechanisms and their timescales (e.g. \citealt{Koopmann2006,Bamford2007,Abramson2011a,Jaffe2011,Cortese2012,Bosch2013,Bretherton2013,Vulcani2016,Schaefer2017,Finn2018,Fossati2018,Vaughan2020}). Just like in 3D-HST \citep{Nelson2015}, which forms our coeval field sample, we use spatially resolved H$\upalpha$ maps of these galaxies obtained using the WFC$3$ G$141$ grism on board the HST and WFC$3$ F$140$W imaging of the same galaxies to accomplish this.

This paper is organized as follows. In Section~\ref{data} we describe the data we use, relevant details regarding grism spectroscopy, the construction of H$\upalpha$ maps and calculation of star formation rates. A summary of our public data release (available at \url{https://www.gclasshst.com}) of all the processed grism data for the 10 clusters this study is based on is provided in Section~\ref{public_data_release} with further details provided in Appendix~\ref{appendix_release}. Section~\ref{methodology} describes the sample selection, our methodology for stacking the data and the size determination process. We present our results in Section~\ref{results}, discussing their physical implications in Section~\ref{discussion}. Finally, we summarise our results in Section~\ref{summary}.

All magnitudes quoted are in the AB system, half-light radii are not circularized and we assume a $\Lambda$CDM cosmology with $\Omega_{m}=0.307$, $\Omega_{\Lambda}=0.693$ and $H_{0}=67.7$~kms$^{-1}$~Mpc$^{-1}$ \citep{Planck2015}.

\section{Data}
\label{data}

\begin{table*}
	\centering
	\caption{The ten GCLASS clusters used in this study. For full names of the clusters, we refer to \protect\cite{Muzzin2012}. $R_{200}$ is the radius at which the mean interior density is 200 times the critical density of the Universe. $M_{200}$ is the mass enclosed within this radius. $\sigma_{v}$ is the line-of-sight velocity dispersion (see \protect\citealt{Biviano2016}). The numbers listed in the last three columns indicate total numbers {\it before} / {\it after} quality checks to \protect\cite{Matharu2018} were applied. Mass completeness limits (see Section~\ref{mass_completeness}) are not applied to these sample numbers.}
	\label{tab:example_table}
	\begin{tabular}{lccccccccc} 
		\hline
		Name & $z_{spec}$ & $M_{200}$ & $R_{200}$ & $\sigma_{v}$ & Spec-{\it z} Members & Grism-{\it z} & Total Members\\
                 &  & [$10^{14}M_{\odot}$] & [kpc] & [km s$^{-1}$] & in HST FOV & Members & in HST FOV\\
		\hline
		SpARCS-0034 & 0.867 &  $2.0\pm0.8$ & $888\pm110$ & $609_{-66}^{+75}$ & 37/33 & 17/16 & 54/49\\
		SpARCS-0035 & 1.335 &  $5\pm2$ & $977\pm154$ & $941_{-137}^{+159}$ & 22/22 & 26/18 & 48/40\\
		SpARCS-0036 & 0.869 &  $5\pm2$ & $1230\pm129$ & $911_{-90}^{+99}$ & 43/41 & 31/30 & 74/71\\
                SpARCS-0215 & 1.004 &  $3\pm1$ & $953\pm103$ & $758_{-77}^{+85}$ & 42/40 & 32/28 & 74/68\\
		SpARCS-1047 & 0.956 &  $3\pm1$ & $926\pm138$ & $680_{-86}^{+98}$ & 22/19 & 15/14 & 37/33\\
		SpARCS-1051 & 1.035 &  $1.2\pm0.5$ & $705\pm102$ & $530_{-65}^{+73}$ & 34/32 & 11/10 & 45/42\\
		SpARCS-1613 & 0.871 &  $13\pm3$ & $1663\pm130$ & $1232_{-93}^{+100}$ & 73/65 & 41/25 & 114/90\\
                SpARCS-1616 & 1.156 &  $3\pm1$ & $854\pm107$ & $701_{-73}^{+81}$ & 38/35 & 19/17 & 57/52\\
		SpARCS-1634 & 1.177 &  $4\pm2$ & $1008\pm131$ & $835_{-82}^{+91}$ & 38/34 & 15/15 & 53/49\\
                SpARCS-1638 & 1.196 &  $1.9\pm0.9$ & $769\pm117$ & $585_{-65}^{+73}$ & 26/23 & 13/9 & 39/32\\
		\hline
	\end{tabular}
\end{table*}

\subsection{The GCLASS Survey}
\label{GCLASS_survey}

The Gemini Cluster Astrophysics Spectroscopic Survey (GCLASS, see \citealt{Muzzin2012} and \citealt{VanderBurg2013}) was the spectroscopic follow-up of $10$ massive clusters in the redshift range \mbox{$0.86<z<1.34$}, drawn from the $42$ deg$^{2}$ Spitzer Adaptation of the Red-sequence Cluster Survey (SpARCS, see \citealt{Muzzin2009}, \citealt{Wilson2009a} and \citealt{Demarco2010}). Extensive optical spectroscopy was obtained with the Gemini Multi-Object Spectrographs (GMOS) on both Gemini-South and -North. In total, $1282$ galaxies obtained a spectroscopic redshift, with $457$ of these being confirmed as cluster members. There is also $12$-band photometry available for these clusters. The details of the photometry in 11 bands ({\it ugrizJK$_{s}$}, $3.6\mu$m, $4.5\mu$m, $5.8\mu$m and $8.0\mu$m) is summarised in Appendix A of \cite{VanderBurg2013}. Newly acquired photometry in the F$140$W band (wide {\it JH} gap) has been obtained as part of the GCLASS HST follow-up (see Section~\ref{HST_data}). The public release of all GCLASS data products are now available, and detailed in \cite{Balogh2020a}. Physical properties of the $10$ clusters in the GCLASS survey are summarised in Table~\ref{tab:example_table}.

\subsection{HST/WFC3 G141 Grism Spectroscopy}
\label{grism_spectroscopy}

A grism is the combination of a diffraction grating and prism. In the context of the G$141$ grism, a series of ridges on the glass surface of a prism act as the diffraction grating, dispersing light from an object at an angle. The prism directs the dispersed light towards the WFC$3$ detector, keeping the light at some chosen central wavelength undeviated. The result is an image of the galaxy at different wavelengths, in $46.5${\AA} increments spanning \mbox{$10750<~\lambda~/$~{\AA}~$<17000$}. An unresolved emission line in this two-dimensional spectrum manifests itself as an image of the galaxy in that line on top of the underlying continuum. The high spatial resolution of the WFC$3$ (FWHM$\sim0.141^{\prime\prime}$ at 14000{\AA}) and the low spectral resolution of the G$141$ grism ({\it R}~$\sim130$) allow for spatially-resolved ($0.130\leqslant\textrm{FWHM~/~arcsec}\leqslant0.156"$ for $11000\leqslant~\lambda~/$~{\AA}~$\leqslant17000$) emission line maps \citep{Nelson2015}.

The wavelength coverage of the G$141$ grism is such that H$\upalpha$ emission can be detected in galaxies within the redshift range \mbox{$0.7<z<1.5$}. This redshift range coincides perfectly with the redshift range of the GCLASS clusters (\mbox{$0.86<z<1.34$}). One spectral resolution element for a galaxy at $z\sim1$ corresponds to a velocity range of approximately \mbox{$1000$ kms$^{-1}$}. The majority of galaxies have line widths well below this. Therefore, structure in the emission line maps is due to spatial morphology, not kinematics \citep{Nelson2015}.

\subsection{GCLASS HST data}
\label{HST_data}

Designed deliberately analogous to 3D-HST observations, the HST follow-up to GCLASS (GO-13845; PI: Muzzin) provides WFC$3$ F$140$W imaging and G$141$ grism spectroscopy for all of the clusters to a $2$-orbit depth. $90\%$ of this time is spent on the grism spectroscopy (see Section~\ref{grism_spectroscopy}) and the remaining $10\%$ on F$140$W imaging. This was done to ensure we could obtain spatially-resolved H$\upalpha$ maps of the star-forming cluster galaxies with a sufficient signal-to-noise ratio, such that stacks containing $\sim30 - 40$ galaxies could have their H$\upalpha$ profiles measured to $\sim10\%$ accuracy \citep{Nelson2013a}. The exposure time of the F140W imaging for 9 of the 10 clusters is $\sim800$~seconds, with a 5$\sigma$ F140W limiting magnitude for galaxies of $\sim26.6$. The exception to this is SpARCS-0035, which has a varying depth, ranging from $\sim800$ to $\sim7460$ seconds\footnote{This cluster was part of The Supernova Cosmology Project ``See Change" program, details of which can be found in \cite{Hayden2021}. As a result, this cluster's observations were deeper, with pointings of different orientations overlapping each other.\label{seechange}}. The G141 grism spectroscopy exposure times for the GCLASS clusters range from 4312 seconds (SpARCS-0215) to 5312 seconds (SpARCS-1616) per pointing.


The HST observations for 9 of the 10 clusters are taken with a WFC3 2-pointing mosaic with a random orientation. For the 10th cluster, SpARCS-1047 (see Table~\ref{tab:example_table}), a single pointing is obtained covering the centre of the cluster. The observations cover approximately $\sim~8$~arcmin$^{2}$ for each cluster. This corresponds to approximately a quarter of the observing area covered by GMOS in the GCLASS survey (Section~\ref{GCLASS_survey}). Four of the GCLASS clusters (SpARCS-0035, SpARCS-1616, SpARCS-1634 and SpARCS-1638; See Table~\ref{tab:example_table}) have imaging going out to $\sim~R_{200}$. Despite this smaller overlap with GCLASS, spectroscopic density is highest in the cores of clusters. $86\%$ of galaxies confirmed as cluster members in GCLASS are in the HST fields of view. Furthermore, grism redshifts derived from the G141 grism spectroscopy (Section~\ref{grism_spectroscopy}) allow for the identification of an additional 182 cluster members, increasing the number of cluster members per cluster by an average of 53\% of the GCLASS spectroscopic sample \citep{Matharu2018}. Many of these cluster members were identified using the H$\upalpha$ emission line and therefore result in a much more complete sample of H$\upalpha$ emitters than from the spectroscopic sample alone. A breakdown of the number of cluster members for each cluster can be seen in Table~\ref{tab:example_table}. 

\subsection{Grism data reduction}
\label{grism_reduction}
The {\it Grism redshift \& line analysis software for space-based slitless spectroscopy} (\texttt{Grizli}\footnote{\url{https://grizli.readthedocs.io/en/master/}}, \citealt{2019ascl.soft05001B}) is used to reduce the G$141$ data and create the H$\upalpha$ line maps. \texttt{Grizli} is a software that allows for a full processing of space-based slitless spectroscopy datasets.

A detailed description of how \texttt{Grizli} works can be found in \cite{Simons2020a}. Here, we summarize the details relevant to the study presented in this paper. \texttt{Grizli} uses the HST proposal ID (GO-13845; PI: Muzzin) to retrieve the data from the MAST archive. Routines detailed in \cite{Gonzaga2012,Brammer2015,Brammer2016} and \cite{Momcheva2016} are used to process the WFC3 F140W imaging and G141 grism data for a variable sky background, hot pixels, cosmic rays, flat-fielding and sky subtraction. 

A full-field contamination model for each 2D grism exposure in each pointing is created to subtract contamination from overlapping spectra of adjacent objects. This contamination model is created by modelling the HST G141 mosaic for each cluster in an iterative process. The first pass creates a contamination model for all sources in the mosaic with F140W magnitude~$ < 25$. This is done by taking the segmentation map (Figure~\ref{fig:data_release}, Panel A) for each source, assuming that the continuum spectra are normalized at F140W and flat in units of flux density, $f_{\lambda}$.  Then a second pass contamination model for all sources with F140W magnitude~$ < 24$ is created by subtracting the contamination model from the first pass and then fitting a third-order polynomial to the spectrum of each source, from brightest to faintest.  The resulting static contamination model of all neighboring sources is subtracted from the 2D spectrum of a given object of interest.

\texttt{Grizli} is used to extract and fit all sources in the GCLASS HST field-of-view with signal-to-noise ratio \mbox{(SNR) > 7} in F140W photometry and with 
F140W magnitude < 26. The number of sources extracted per cluster ranges from 573 (SpARCS-1047) to 1484 (SpARCS-0035). For each source, the redshift is determined by simultaneously fitting the 12-band multiwavelength photometry (see Section~\ref{GCLASS_survey}) and the G141 grism spectrum. This is done by including the photometric redshift p($z$) as a prior on the grism spectrum redshift fit that is multiplied by the spectroscopic redshift p($z$)\footnote{The current version of \texttt{Grizli} uses an improved method for fitting the photometry + grism spectrum simultaneously. See ``Improvement in the grism/photometry scaling algorithm" in the \texttt{Grizli} documentation for more details.}.

\subsubsection{Grism redshift determination}
\label{grism_z}
Grism redshifts are determined by using a basis set of template Flexible Stellar Population Synthesis models (\texttt{FSPS}; \citealt{Conroy2009,Conroy2010}) that are projected to the pixel grid of the 2D grism exposures using the spatial morphology from the F140W image. The 2D template spectra are then fit to the observed spectra with non-negative least squares. When performing a redshift fit, the user provides a trial redshift range. For the GCLASS extraction, $0<z<5$ was used. The final grism redshift is taken to be where the  posterior, including the (optional) photometric redshift prior, is maximized across the range of trial redshifts provided by the user.

\subsection{Making H\texorpdfstring{$\alpha$}{TEXT} maps}
\label{making_maps}

As part of the grism redshift determination process (see Section~\ref{grism_z}), a combination of continuum templates and line complex templates are used. The line complex templates include [\ion{O}{2}] + [\ion{Ne}{3}], [\ion{O}{3}] + H$\upbeta$ and H$\upalpha$ + [\ion{S}{2}]. As well as breaking redshift degeneracies, the redshift determination process leads to the generation of a full continuum + line model for each spectrum in 2D. This then allows the user to create drizzled\footnote{going from the distorted to undistored mosaic.} continuum-subtracted narrow-band maps at any desired output wavelength. Emission line maps are therefore created by deliberately choosing the wavelength of the detected emission line from the redshift determination process (e.g. $\uplambda = 1.245\upmu$m for H$\upalpha$ in Figure~\ref{fig:data_release}). The full World Coordinate System (WCS) information for each individual grism exposure of the source is used.

Emission line maps are generated by subtracting the best-fit continuum model with the assumption that the direct image (Figure~\ref{fig:data_release}, Panel A) represents the morphology of the source in the continuum. For GCLASS, these emission line maps are generated with a pixel scale of 0.1$^{\prime\prime}$ and dimensions of $80\times80$ pixels. Example emission line maps for a galaxy in the final cluster sample of our study can be seen in Panel D of Figure~\ref{fig:data_release}.

\subsubsection{Cleaning H\texorpdfstring{$\alpha$}{TEXT} maps}
\label{processing_maps}
There are some cases in which the contamination subtraction perfomed by \texttt{Grizli} (see Section~\ref{grism_reduction}) is sub-optimal, leaving traces of contamination in the resulting emission line maps. In such cases, we place those H$\upalpha$ maps into two categories: those that can have all their contamination masked using the \texttt{Grizli}-generated contamination map thumbnail and those that will require additional masking which the contamination map thumbnail is unable to fully account for.

For the first category, we mask pixels in the emission line map where the contamination exceeds \mbox{$0.03\times10^{-17}$~ergs~s$^{-1}$~cm$^{-2}$}. This threshold was found to work well in cleaning the H$\upalpha$ maps of our comparative field study \citep{Nelson2015}. For the second category, we create masks based on the specific requirements of each map since the \texttt{Grizli}-generated contamination map does not include all the contaminated pixels. In other words, we mask additional pixels that are not accounted for in the \texttt{Grizli}-generated contamination map.

\subsection{Integrated star formation rates}
\label{sfrs}
The star formation rates (SFRs) used in our study are calculated from total H$\upalpha$ fluxes determined by \texttt{Grizli} using the same method that was used for the calculation of H$\upalpha$-derived SFRs from the 3D-HST G141 grism spectroscopy in \cite{Whitaker2014a}.

As in \cite{Nelson2015}, to account for the blending of H$\upalpha$ and [N~{\small II}], the H$\upalpha$ flux is scaled down by the normalization factor $1+f(\textrm{[N {\small II}]/H}\upalpha)$ before SFRs are calculated. The rescaled H$\upalpha$ flux, H$\upalpha_{\textrm{rescaled}}$, is defined as:

\begin{equation}
\label{Ha_rescaled}
\textrm{H}\upalpha_{\textrm{rescaled}}=\frac{\textrm{H}\upalpha_{\textrm{measured}}}{1+f(\textrm{[\ion{N}{2}]}/\textrm{H}\upalpha)}
\end{equation}

When calculating SFRs for individual galaxies (see Figure~\ref{fig:SFMS}), a fixed value of 0.15 was used for $f(\textrm{[\ion{N}{2}]/H}\upalpha)$. For the H$\upalpha$ stacks, the same relation that was used in \cite{Nelson2015} from \cite{Matthee2015} shown below is used to obtain the flux ratio between the [\ion{N}{2}] and H$\upalpha$ emission, $f(\textrm{[\ion{N}{2}]/H}\upalpha)$, for each stack:

\begin{equation}
\label{Ha_rescaling}
f(\textrm{[\ion{N}{2}]/H}\upalpha)= -0.296\times\textrm{Log}_{10}(\textrm{EW}_{\textrm{H}\upalpha+\textrm{[N II]}})+0.8
\end{equation}

For the equivalent width of the H$\upalpha+$[\ion{N}{2}] emission, EW$_{\textrm{H}\upalpha+\textrm{[N II]}}$, we use the H$\upalpha$ equivalent widths calculated by \texttt{Grizli} for each galaxy.

For the measurements shown in Figure~\ref{fig:SFMS}, after rescaling the H$\upalpha$ flux for the [\ion{N}{2}] contribution, we calculate SFRs using the H$\upalpha$ flux conversion from \cite{Kennicutt1998}, but adapting it from a Salpeter initial mass function (IMF) to a \cite{Chabrier2003} IMF following the method in \cite{Muzzin2010} shown below in Equation~\ref{eq:sfr_muzzin}:

\begin{equation}
\label{eq:sfr_muzzin}
\frac{\mathrm{SFR(H}\upalpha\mathrm{)}}{\mathrm{M}_{\odot}\mathrm{yr}^{-1}} = \frac{1.7\times10^{-8} L_{\mathrm{H}\upalpha}}{\mathrm{L}_{\odot}}
\end{equation}

\subsection{Stellar masses}

Stellar masses are estimated using \texttt{FAST} \citep{Kriek2009} with the redshift of each cluster galaxy fixed to the spectroscopic redshift of its cluster's center. These stellar masses are then corrected for the difference between the total F140W flux from the photometric catalog and the total F140W flux as measured by \texttt{GALFIT} (see Section~\ref{size_determination}). An exponentially declining star formation history parameterization is assumed. Further details on the calculation of stellar masses can be found in \cite{Matharu2018}.

\section{Public Data Release}
\label{public_data_release}

\begin{figure*}

	\centering\includegraphics[width=\textwidth]{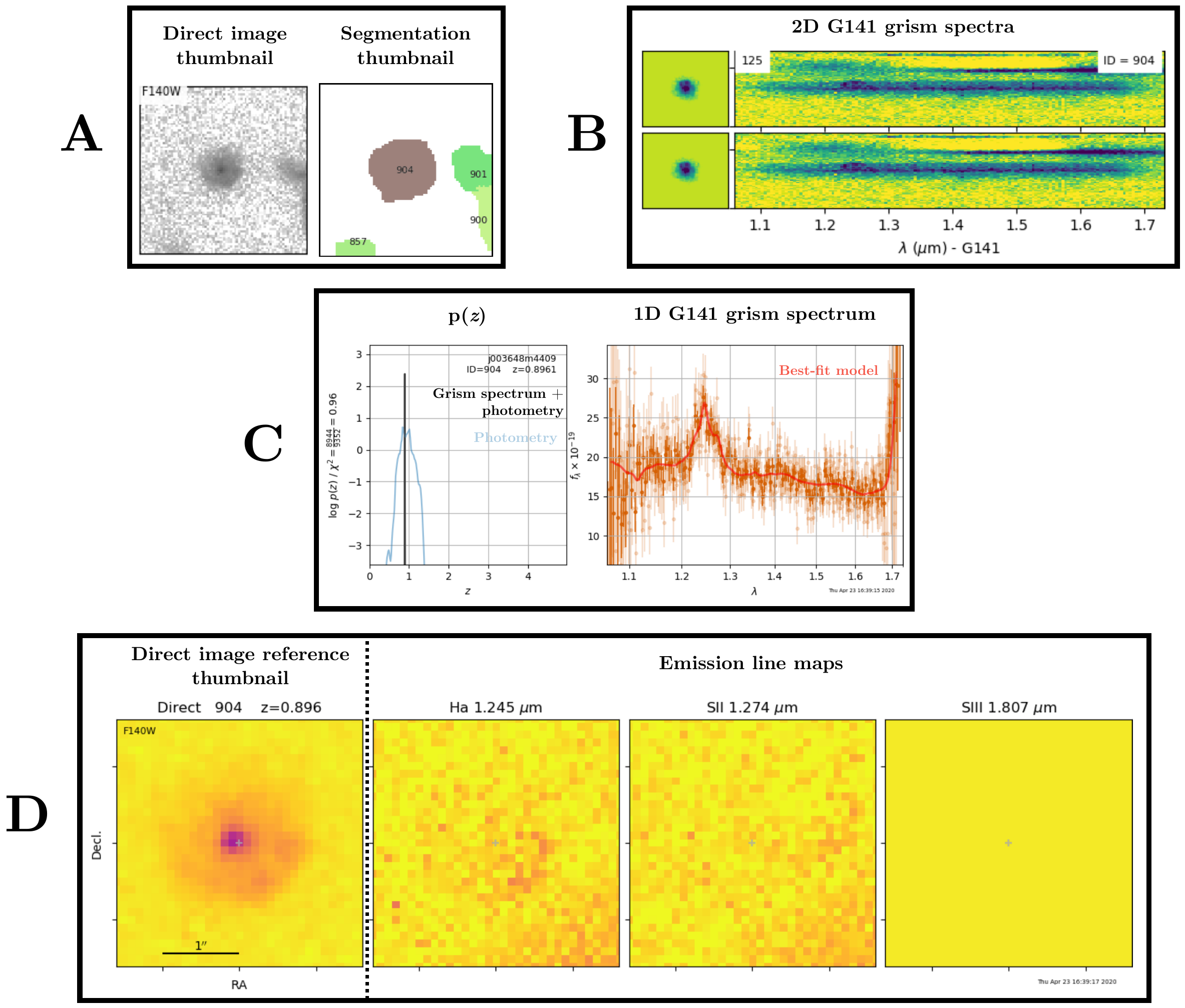}
    \caption{Summary of data products output by \texttt{Grizli} for a galaxy in the cluster sample. Panel A shows the drizzled $80\times80$ pixel, 0.1$^{\prime\prime}$ pixel scale F140W direct image thumbnail, along with the segmentation map for the thumbnail. Individual segments are labelled by their source ID. In the top row of panel B, a small thumbnail of the galaxy in F140W with the raw two-dimensional (2D) G141 grism spectrum is shown to its right. The number in the top left of the grism spectrum is the position angle (PA) of the G141 grism. The bottom row shows the PA-combined 2D grism spectrum. For GCLASS, only one PA is available, so the PA-combined grism spectrum is the same as the single PA grism spectrum shown above it. The first plot in panel C shows the p($z$) from fitting the photometry alone (blue line) and fitting the grism spectrum + photometry simultaneously (black line). For some galaxies in the data release, for which a spectroscopic redshift is available, an upside down red triangle pointing towards the location of the spectroscopic redshift on the $x$-axis will also be visible. The spectroscopic redshift in red font will also appear in the top right-hand corner. The unique field identifier (see website), source ID and grism redshift is stated in the upper right-hand corner of this plot. The second plot in panel C shows the one-dimensional (1D) grism spectrum (orange data points with error bars) with the best-fit model (solid red line) over-plotted. The first panel in panel D shows a zoomed-in region of the blotted (going from the undistorted to the distorted image, see website) direct image thumbnail. The source ID and grism redshift are provided in its title. The second panel in panel D shows the emission line maps for this galaxy, with the abbreviation of the emission line and its observed wavelength in the title of each map.}
    \label{fig:data_release}
\end{figure*}

We make available all the \texttt{Grizli} data products for the GCLASS HST data at the following website: \mbox{\url{https://www.gclasshst.com}}. In this section, we summarize the main data products such that users can more easily navigate them according to their needs. Further details on all data products can be found in Appendix~\ref{appendix_release} and at the data release website. For all sources with SNR > 7 in F140W photometry and with F140W magnitude < 26, the data products released include:

\begin{itemize}
    \item F140W direct image thumbnails (Section~\ref{direct_image_thumbnails} and Panel A, Figure~\ref{fig:data_release})
    \item Two-dimensional (2D) G141 grism spectra for each observation of a source (Section~\ref{2dspec}).
    \item Stacked 2D G141 grism spectra for each position angle (PA) of the grism the source is observed at (Section~\ref{2dspec} and top row of Panel B, Figure~\ref{fig:data_release}).
    \item PA-combined 2D G141 grism spectrum for each source. (Section~\ref{2dspec} and bottom row of Panel B, Figure~\ref{fig:data_release}).
    \item One-dimensional (1D) G141 grism spectrum for each source with best-fit Spectral Energy Distribution (Section~\ref{1dspec} and right-hand plot of Panel C, Figure~\ref{fig:data_release}).
    \item $p(z)$ distributions and parameters for the $z_{grism}$ fit (Section~\ref{emaps} and left-hand plot of Panel C, Figure~\ref{fig:data_release}).
    \item Emission line maps for each source (Section~\ref{emaps} and right-hand side of Panel D, Figure~\ref{fig:data_release}).
\end{itemize}

\section{Methodology}
\label{methodology}

\subsection{Sample selection}
\label{sample_selection}

\begin{table}
	\centering
	\caption{Summary of Sample Selection. $R_{\mathrm{eff, F140W}}$ and $[b/a]_{\mathrm{F140W}}$ are the GALFIT half-light radius and axis ratio measured from the HST WFC3 F140W imaging, respectively.}
	\label{tab:sample_table}
	\begin{tabular}{l} 
		\hline
		Sample Selection Criteria \\
		\hline
		$0.86<z<1.34$\\
		spec-$z$ or grism-$z$ secure cluster member \\

		F140W magnitude $< 25$ \\
		$R_{\mathrm{eff, F140W}} < 50$ kpc\\
		Log$(M_{*}/\mathrm{M}_{\odot}) \geqslant 9.60$ \\
        
        $\begin{cases}
        
		(U-V)_{\mathrm{rest}} < 1.44 & (V-J)_{\mathrm{rest}} < 0.88\\
		(U-V)_{\mathrm{rest}} < 0.81~(V-J)_{\mathrm{rest}}+0.73 & 0.88 < (V-J)_{\mathrm{rest}} < 1.57\\
		(V-J)_{\mathrm{rest}} > 1.57 & \text{otherwise}
		
        \end{cases}$\\
        
		H$\upalpha$ SNR $> 3$ \\

		$[b/a]_{\mathrm{F140W}} \geqslant 0.3$ \\

		\hline
	\end{tabular}
\end{table}

This section will explain how the sample selection was carried out for this study. However, a summary of the sample selection can be viewed in Table~\ref{tab:sample_table}.

\subsubsection{Spec-\texorpdfstring{$z$}{TEXT} and Grism-\texorpdfstring{$z$}{TEXT} cluster member selection}
\label{z_members}

Our first step in sample selection is to select all the secure spec-$z$ cluster members identified as part of the GCLASS survey that are in the HST fields-of-view (see Section 4.1 of \citealt{Muzzin2012} for details on the cluster membership criteria). This amounts to 375 cluster galaxies in total (quiescent and star-forming) across all 10 GCLASS clusters (6th column, Table~\ref{tab:example_table}, {\it before} quality check).

The G141 grism data for all extracted sources in the GCLASS HST field-of-view are quality-checked by eye. The criteria for this quality check is detailed in Section 2.1.4 of \cite{Matharu2018}. Galaxies which obtained a good quality spectroscopic redshift as part of GCLASS and good quality G141 grism data were used to determine a ``secure cluster" selection threshold on grism redshifts. This ``secure cluster" selection threshold allowed for additional cluster members that do not have GMOS spectroscopy from GCLASS, but do have well-determined grism redshifts to be identified. This selection threshold was:
\begin{equation}
\label{eq:threshold}
-0.02<\frac{z_{grism} - z_{cluster}}{1+z_{grism}}<0.02
\end{equation}
where $z_{grism}$ is the \texttt{Grizli}-determined grism redshift and $z_{cluster}$ is the spectroscopic redshift of the cluster center\footnote{Which in most cases, is the redshift of the Brightest Cluster Galaxy (BCG).}. We refer the reader to Section 3 of \cite{Matharu2018} for details on how this secure cluster selection threshold was determined. Cluster members that were flagged as ``low confidence" in GCLASS were checked for good quality grism redshifts consistent with their respective cluster redshifts. This process adds 220 grism-$z$ secure cluster members (7th column, Table~\ref{tab:example_table}, {\it before} quality check) to the original 375 identified from GCLASS slit spectroscopy. This amounts to a total of 595 cluster galaxies (quiescent and star-forming).

\subsubsection{Sample selections based on the size determination process}
\label{GALFIT_selection}

We then apply quality checks on the GALFIT results in F140W for these galaxies, detailed in Section 2.3.4 of \cite{Matharu2018}. Then we determine a reliability threshold on size measurements by running our size determination method developed in \cite{Matharu2018} for the F140W imaging against that of \cite{VanderWel2012} by comparing our results for the same set of galaxies that are at $0.86<z<1.34$ in the F160W CANDELS-COSMOS mosaic. We found that those galaxies with GALFIT measurements of $R_{\mathrm{eff}}<50$~kpc and F160W magnitude $<25$ exhibited the best agreement in $R_{\mathrm{eff}}$ measurements (0.28\% mean offset towards smaller sizes for our size determination method) between the two methods. After applying these GALFIT-related sample selections, our cluster sample is reduced to 344 spec-$z$ members and 182 grism-$z$ members (6th and 7th column, Table~\ref{tab:example_table}, {\it after} quality check). Leading to a total of 526 cluster galaxies (quiescent and star-forming).

\subsubsection{Mass completeness limits}
\label{mass_completeness}

The mass completeness limits for the final cluster sample are set to match the grism spectroscopic completeness limits of the final cluster sample, and are calculated for star-forming and quiescent galaxies separately (see Section~\ref{colors} for details on the star-forming classification). Every cluster galaxy in the final sample of 526 cluster galaxies (spec-$z$ and grism-$z$ cluster member, reliable grism spectra, F140W magnitude~$<25$ and $R_{\mathrm{eff, F140W}}<50$~kpc) has been selected such that it has a good quality grism spectrum from which reliable grism redshifts can be calculated, confirming its cluster membership.

Within the star-forming and quiescent cluster samples, the cluster galaxy with the highest mass-to-light ratio is found. Then the faintest cluster galaxy in the sample is found. The stellar mass for this galaxy is calculated for the case where it has a mass-to-light ratio equal to the aforementioned galaxy. The result is the lowest stellar mass for which a reliable, good quality grism spectrum can be obtained in our final cluster sample. We find the star-forming and quiescent mass completeness limits are  Log$(M_{*}/\mathrm{M}_{\odot})=9.60$ and Log$(M_{*}/\mathrm{M}_{\odot})=9.96$ respectively \citep{Matharu2018}. Similar mass completeness limits were found for GCLASS using ground-based {\it K}-band data in \cite{VanderBurg2013}. The increased depth of the F140W imaging allows us to reach slightly lower stellar masses than \cite{VanderBurg2013}. Applying these mass completeness limits to the final cluster sample brings the sample down from 526 to 474 (quiescent and star-forming) cluster galaxies.

\subsubsection{Rest-frame colors}
\label{colors}

$U-V$ and $V-J$ rest-frame colors for each galaxy are obtained using \texttt{EAZY} \citep{Brammer2008}. Further details can be found in \citep{VanderBurg2013}. These rest-frame colors are used to determine the $UVJ$ color separation for star-forming and quiescent cluster galaxies in the sample. The $UVJ$ color separation technique has proved to be successful in separating star-forming and quiescent populations, even when the former is reddened by dust \citep{Wuyts2007, Williams2008, Patel2012a, Foltz2015}. The star-forming GCLASS $UVJ$ selection used is:

\begin{equation}
\label{eq:sf}
\begin{cases}
(U-V)_{\mathrm{rest}} < 1.44 & (V-J)_{\mathrm{rest}} < 0.88\\
(U-V)_{\mathrm{rest}} < 0.81~(V-J)_{\mathrm{rest}}+0.73 & 0.88 < (V-J)_{\mathrm{rest}} < 1.57\\
(V-J)_{\mathrm{rest}} > 1.57 & \text{otherwise}
\end{cases}
\end{equation}

After applying this selection to the sample of 474 cluster galaxies selected so far, we are left with 177 star-forming cluster galaxies.

\subsubsection{H\texorpdfstring{$\alpha$}{TEXT} map quality check}

Out of the 177 star-forming cluster galaxies selected thus far, the \texttt{Grizli} data reduction process (Section~\ref{grism_reduction}) led to 140 of them having H$\upalpha$ maps with H$\upalpha$ flux > 0 (i.e, 37 $UVJ$-selected star-forming cluster galaxies are not detected in H$\upalpha$ at the depth of our observations). We then quality-check these H$\upalpha$ maps and remove the following:
\begin{itemize}
    \item Defective H$\upalpha$ maps due to interlacing problems as a result of the data reduction process.
    \item Possible Active Galactic Nuclei\footnote{Broad-line AGN activity unrelated to star formation manifests itself as stretched out H$\upalpha$ morphologies in the spectral direction \citep{Nelson2015}.} (AGN).
    \item Interacting galaxies\footnote{Interacting or merging galaxies complicate the interpretation of the final stacks.}.
    \item H$\upalpha$ maps with poorly subtracted contamination from nearby sources.
    \item H$\upalpha$ maps with poorly subtracted continuum from the target galaxy itself.
\end{itemize}

After this quality-check, we are left with 85 star-forming cluster galaxies with good quality H$\upalpha$ emission line maps.

\subsubsection{Cuts on H\texorpdfstring{$\alpha$}{TEXT} SNR and F140W axis ratio}

We further remove H$\upalpha$ maps with low integrated SNRs~$\leqslant3$ which contribute mostly noise to the final H$\upalpha$ stacks (see Section~\ref{stacking} for details on the stacking process). We note here that a H$\upalpha$ SNR selection was not used in the field environment study we compare to \citep{Nelson2015}. However, given our much smaller sample (10s of galaxies per stack instead of 100s), low H$\upalpha$ SNR maps hamper our ability to make a reliable size measurement with GALFIT on the final stack.

Finally, to reduce the effects of dust extinction, we remove edge-on galaxies as determined from their F140W axis ratios, $[b/a]_{\mathrm{F140W}}$. Galaxies with $[b/a]_{\mathrm{F140W}}<0.3$ as determined from their individual GALFIT results in \cite{Matharu2018} are removed from the sample. This leaves us with a sample of 65 star-forming cluster galaxies with good quality H$\upalpha$ maps. Some works have found evidence that the level of dust extinction is different for the stellar continuum and H$\upalpha$ emission (e.g. \citealt{Calzetti1999,ForsterSchreiber2009,Munoz-Mateos2009,Yoshikawa2010,Mancini2011,Wuyts2011,Wuyts2013, Kashino2013, Wilman2020, Greener2020}). However, there is no evidence that the relative effects of dust on the stellar continuum and H$\upalpha$ emission change with environment at $z\sim1$. This is perhaps unsurprising at high redshift ($z\geqslant1$) where the majority of studies have shown that the star-forming cluster galaxies are likely to have been recently accreted from the field. Indeed, high redshift star-forming cluster galaxies have been shown to have similar specific star formation rates (SSFRs, e.g. \citealt{Koyama2013,Tran2016a,Nantais2020}), stellar mass functions (SMFs, e.g. \citealt{VanderBurg2013,VanderBurg2020}), metallicities (e.g. \citealt{Kacprzak2015,Tran2015}), kinematics \citep{Alcorn2016}, sizes and morphologies \citep{Tran2016a,Matharu2018} to co-eval field star-forming galaxies. Furthermore, most studies in the literature argue for rapid environmental quenching in high redshift galaxy clusters (see \citealt{Foltz2018} and references therein). Therefore, catching a star-forming cluster galaxy in the process of environmentally quenching is rare at high redshift (see also Section~\ref{outsidein} for a detailed discussion). It is more likely that the majority of star-forming cluster galaxies observed at high redshift have only recently joined the cluster environment. Whilst dust is certainly present in cluster star-forming galaxies, the current body of evidence suggests there is no differential effect of dust between cluster and field environments. Since the analysis in our study focuses on the differential measurement of $R[\mathrm{H}\upalpha/\mathrm{C}]$ between cluster and field star-forming galaxies, it seems reasonable to assume that the effects of dust on our results are negligible.

\subsection{The star formation main sequence of the sample}
\begin{figure*}

	\centering\includegraphics[width=0.8\textwidth]{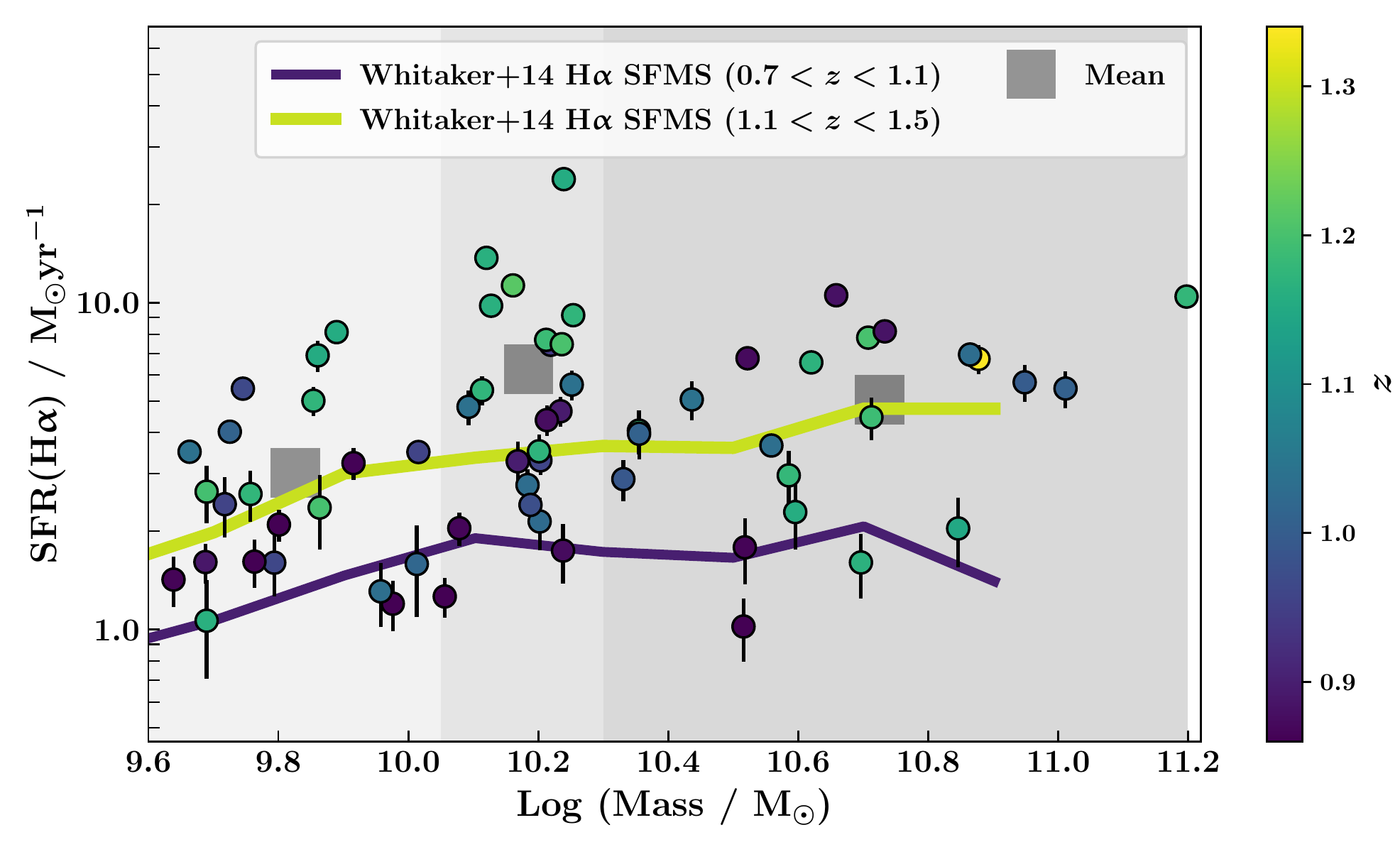}
    \caption{The star formation main sequence (SFMS) of the cluster sample. The grey shaded regions in the background delineate the stellar mass bins for each stack. The large square markers indicate the mean SFR for each stack, plotted at the mean stellar mass for each stack. For reference, we over-plot the H$\upalpha$ SFMS for two redshift bins that approximately span the redshift range of our sample from Figure 8 of \cite{Whitaker2014a}. Individual points and the SFMSs from \cite{Whitaker2014a} are color coded by redshift.}
    \label{fig:SFMS}
\end{figure*}

To illustrate the range in redshift and H$\upalpha$ fluxes the final cluster sample spans, we show the star formation main sequence (SFMS) of the sample in Figure~\ref{fig:SFMS}. 

As detailed in Section~\ref{sfrs}, we follow the same method explained in Section 6.2 of \cite{Whitaker2014a} to calculate H$\upalpha$ SFRs from our G141 grism spectroscopy. We also over-plot the H$\upalpha$ SFMSs from  Figure 8 of \cite{Whitaker2014a} that happen to approximately span the redshift range of our sample. Our measurements and the SFMSs from \cite{Whitaker2014a} are colored by redshift as shown by the color bar.

In general, it can be seen that the cluster star-forming galaxies in our sample are fairly typical, with a H$\upalpha$ SFMS following the field H$\upalpha$ SFMS (as also seen by many other works, e.g. \citealt{Koyama2013,Nantais2020}) from 3D-HST which forms our comparative field sample.

\subsection{Stacking}
\label{stacking}

Due to the difficulty in measuring reliable sizes from individual H$\upalpha$ emission line maps of 2-orbit depth from WFC3 G141 grism spectroscopy, we stack our data as in \cite{Nelson2015} to boost SNR, allowing us to trace the H$\upalpha$ distribution out to large radii. It is worth noting that \texttt{Grizli} did not exist at the time the similar study to ours in the coeval field environment, 3D-HST \citep{Nelson2015} was conducted. Due to the sophistication of \texttt{Grizli}, many of the cleaning and masking techniques employed in \cite{Nelson2015} are no longer required or are completed using improved methods in our study. We therefore follow the stacking method outlined in \cite{Nelson2015} as closely as possible, but deviate in places where \texttt{Grizli} provides improved solutions.

The stacking of H$\upalpha$ maps and F140W direct image thumbnails is done in three bins of stellar mass -- $9.60<\mathrm{Log}(M_{*}/\mathrm{M}_{\odot})\leqslant10.05$, $10.05<\mathrm{Log}(M_{*}/\mathrm{M}_{\odot})\leqslant10.3$ and  $10.3<\mathrm{Log}(M_{*}/\mathrm{M}_{\odot})\leqslant11.2$ -- ensuring an approximately equal number of galaxies are in each stack. First, we convert the units of the F140W direct image thumbnails (which are in \mbox{counts s$^{-1}$}) to the same units the H$\upalpha$ maps are in (\mbox{$\times10^{-17}$~ergs~s$^{-1}$~cm$^{-2}$}) using the pivot wavelength of the F140W filter. Bad pixels are masked and then the F140W direct image segmentation thumbnails (see example in Panel A, Figure~\ref{fig:data_release}) are applied to {\it both} the H$\upalpha$ maps and F140W direct image thumbnails to mask neighbouring sources. It is worth noting here that the asymmetric double pacman mask devised in \cite{Nelson2015} is not required for our H$\upalpha$ emission line maps output by \texttt{Grizli}. This is because \texttt{Grizli} incorporates improved techniques for masking neighbouring [\ion{S}{2}] emission and performing stellar continuum subtraction.  Specifically, the flux of the  [\ion{S}{2}]$\lambda\lambda6717,6731$ doublet (assuming a 1:1 line ratio) is determined directly from the spectrum (as for H$\upalpha$ and the other lines) and the 2D [\ion{S}{2}] model, created assuming the same spatial morphology from the F140W thumbnail, is subtracted before drizzling the H$\upalpha$ map.

Each pixel in each H$\upalpha$ map and F140W direct image thumbnail is then weighted by its value in the corresponding weight (inverse variance) map. As in \cite{Nelson2015}, each H$\upalpha$ map and F140W direct image thumbnail is weighted by its F140W flux density\footnote{These are the F140W fluxes of the best-fit \texttt{Grizli} template.} such that no stack is dominated by a single bright galaxy. For each stack, these normalized maps and direct image thumbnails are summed. The corresponding weight maps are also summed for each stack. Exposure-corrected stacks are created by dividing the summed science stacks by the summed weight stacks for each stellar mass bin. Variance maps for each stack are then simply $\sigma_{ij}^2 = 1/\sum{w_{ij}}$, where $w_{ij}$ is the weight map for each object in the stack.

We do not rotate and align the thumbnails along the measured semi-major axis since \cite{Nelson2015} demonstrated that doing so did not change their results.

\begin{figure*}

	\centering\includegraphics[width=\textwidth]{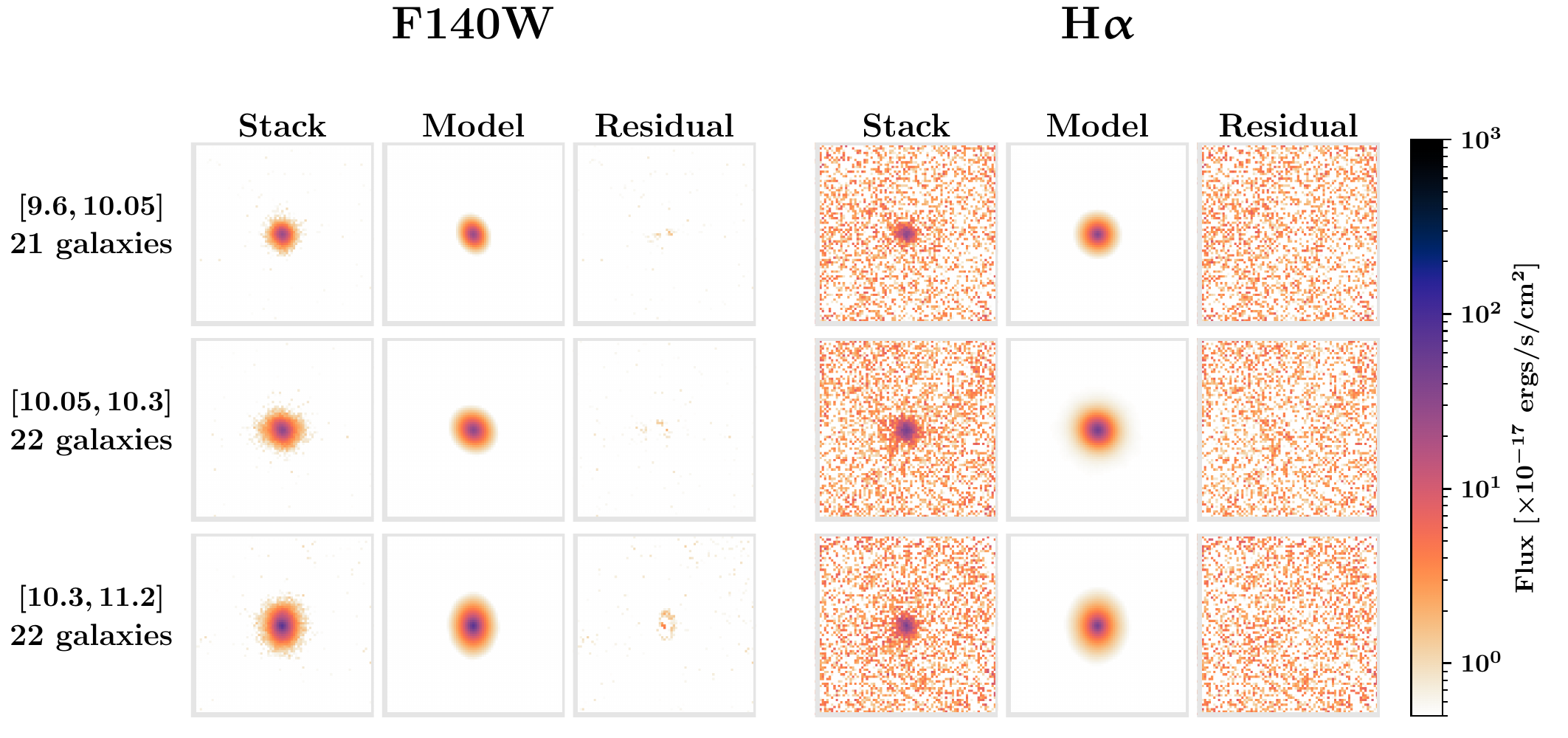}
    \caption{ The stellar continuum and H$\upalpha$ stacks for the cluster sample with their associated GALFIT fits. The numbers in square brackets in the first column state the Log($M_{*}$/M$_{\odot}$) range of each stack. 1 pixel = 0.1$^{\prime\prime}$. The colormap is logarithmic, with H$\upalpha$ stacks and fits multiplied by a factor 100 for visibility.}
    \label{fig:fits}
\end{figure*}

We account for the the blending of H$\upalpha$ and [\ion{N}{2}] in each stack by rescaling the H$\upalpha$ flux using the method described in Section~\ref{sfrs}. The median equivalent width of H$\upalpha$ for each stack is used when doing this. The final F140W and H$\upalpha$ stacks can be seen in Figure~\ref{fig:fits}.

\subsection{Size determination}
\label{size_determination}
The size determination process for measuring the spatial extent of the F$140$W and H$\upalpha$ emission in each stack is conducted using GALFIT \citep{Peng2002b, Peng2010a}, following the two-GALFIT-run approach developed and used in \cite{Matharu2018}.

GALFIT is software that fits two-dimensional analytic functions to light profiles in an image \citep{Peng2002b, Peng2010a}. We fit our stacks with a single-component S\'ersic profile, defined as:

\begin{equation}
    I(r)=I(R_{\mathrm{eff}})\exp\left[-\kappa\left(\left(\frac{r}{R_{\mathrm{eff}}}\right)^{\frac{1}{n}}-1\right)\right]
	\label{eq:sersic_profile}
\end{equation}
where $R_{\mathrm{eff}}$ is the half-light radius. This is the radius within which half of the galaxy's total flux is emitted\footnote{In our study, the half-light radius measurements are {\it not} circularized.}. $n$ is the S\'ersic index and $\kappa$ is an $n$-dependent parameter. $I(R_{\mathrm{eff}})$ is the intensity at the half-light radius.

\subsubsection{PSF construction}

Note that our method for creating point-spread functions (PSFs) for each stack improves upon the method used in the comparative field environment study (3D-HST, \citealt{Nelson2015}), where a single PSF, modelled using the TinyTim software \citep{Krist1995} was used in the GALFIT size determination process for all stacks.

As well as sigma (or noise) images (see Section~\ref{stacking}), GALFIT requires a PSF to account for image smearing due to the resolution limit of WFC$3$. Within \texttt{Grizli}, PSF thumbnails in F$140$W for each galaxy are generated using the WFC$3$/IR empirical PSF library from \cite{Anderson2016}. PSFs exist for various positions across the detector sampled on a $3\times3$ grid. At each of these grid points, four sub-pixel center positions are available. For each galaxy, the relevant empirical PSF is placed at the exact location of the galaxy in the detector frame of each individual exposure within which the galaxy is detected. These PSF models are then drizzled to the same pixel grid as the emission line maps \citep{Mowla2018}. 

We create PSFs for each stack by stacking the F$140$W PSFs for individual galaxies output by \texttt{Grizli}. The PSFs for each stack are summed, then divided by a mask indicating how many individual PSFs in the stack contained flux per pixel. The F$140$W PSF stack is used on both the F$140$W and H$\upalpha$ stack for each stellar mass bin.

\subsubsection{Size measurements with GALFIT}
\label{size_measurements}

We use a two-GALFIT-run approach to measure the half-light radii of each stack, similar to the two-GALFIT-run approach used in \cite{Matharu2018}. For the first GALFIT run on the stacks, we keep all parameters free. For each individual galaxy in our sample, F140W GALFIT measurements were obtained in \cite{Matharu2018}. For each F140W stack, we therefore take the mean values of the GALFIT-determined F140W magnitude, half-light radius, S\'ersic index, axis ratio and position angle (PA) of all galaxies in the stack, and use those as our initial values for GALFIT. Initial values for the $x$ and $y$ pixel coordinates are set to the central pixel of the thumbnail, since each source is centered in all \texttt{Grizli}-generated thumbnails. Since neighbouring sources have been masked in each thumbnail during the stacking process (see Section~\ref{stacking}), we do not fit for adjacent sources in each thumbnail as was done in \cite{Matharu2018}. The purpose of this run is to obtain refined values for the galaxy stack shape parameters (pixel coordinates, axis ratio and PA).

After the initial GALFIT run is complete, we check the quality of the fits and the measurements. All poor quality fits and/or those with unreliable measurements were due to the half-light radius, S\'ersic index and/or axis ratio converging to an unphysical value, or reaching a problematic value, which is placed in asterisks by GALFIT. These fits are re-run by fixing the half-light radius and/or S\'ersic index, depending on which has the unphysical/problematic value. Since we are only interested in obtaining refined values for the shape parameters, fixing parameters which the user is unconcerned about is recommended when GALFIT is struggling to converge to a sensible solution. Once all the F140W stacks have reliable shape parameter measurements, we set up the second GALFIT run. The second GALFIT run uses the refined values for pixel coordinates, axis ratio and PA determined in the first GALFIT run, but keeps them fixed in the fit. The only parameters left free are magnitude, half-light radius and S\'ersic index. We then check the quality of the fits and the measurements after the second run. For all three F140W stacks, the final measurements for half-light radius, S\'ersic index and magnitude converged with good quality fits, as can be seen in Figure~\ref{fig:fits}.

The lower SNR H$\upalpha$ stacks make for a more challenging task for GALFIT to determine reliable size measurements. We therefore adapt our two-GALFIT-run approach using reasonable assumptions. In our initial GALFIT run, we fix the pixel coordinates and the PA to the final values determined by GALFIT for the corresponding F140W stacks. It is reasonable to assume that the centroid and PA of the H$\upalpha$ distribution is coincident with the stellar continuum. The KMOS$^{\mathrm{3D}}$ survey for example, which covers the redshift range of the GCLASS sample, found that only $24\%$ of galaxies had H$\upalpha$ distributions which were mis-aligned with their stellar continuum by more than $30^{\circ}$ \citep{Wisnioski2019}. All other parameters -- magnitude, axis ratio, half-light radius and S\'ersic index -- are kept free in the fit with the same initial values as were used for the corresponding F140W stack. After checking the quality of the fits and measurements, the second GALFIT run then fixes the axis ratio to the value determined in the first run, leaving the magnitude, half-light radius and S\'ersic index free. The resulting fits, along with the H$\upalpha$ stacks, can be seen in Figure~\ref{fig:fits}.

\section{Results}
\label{results}
\subsection{The stellar continuum vs. H\texorpdfstring{$\alpha$}{TEXT} mass--size relation in the cluster and field environments at \texorpdfstring{$z\sim1$}{TEXT}}

\begin{figure*}

	\centering\includegraphics[width=\textwidth]{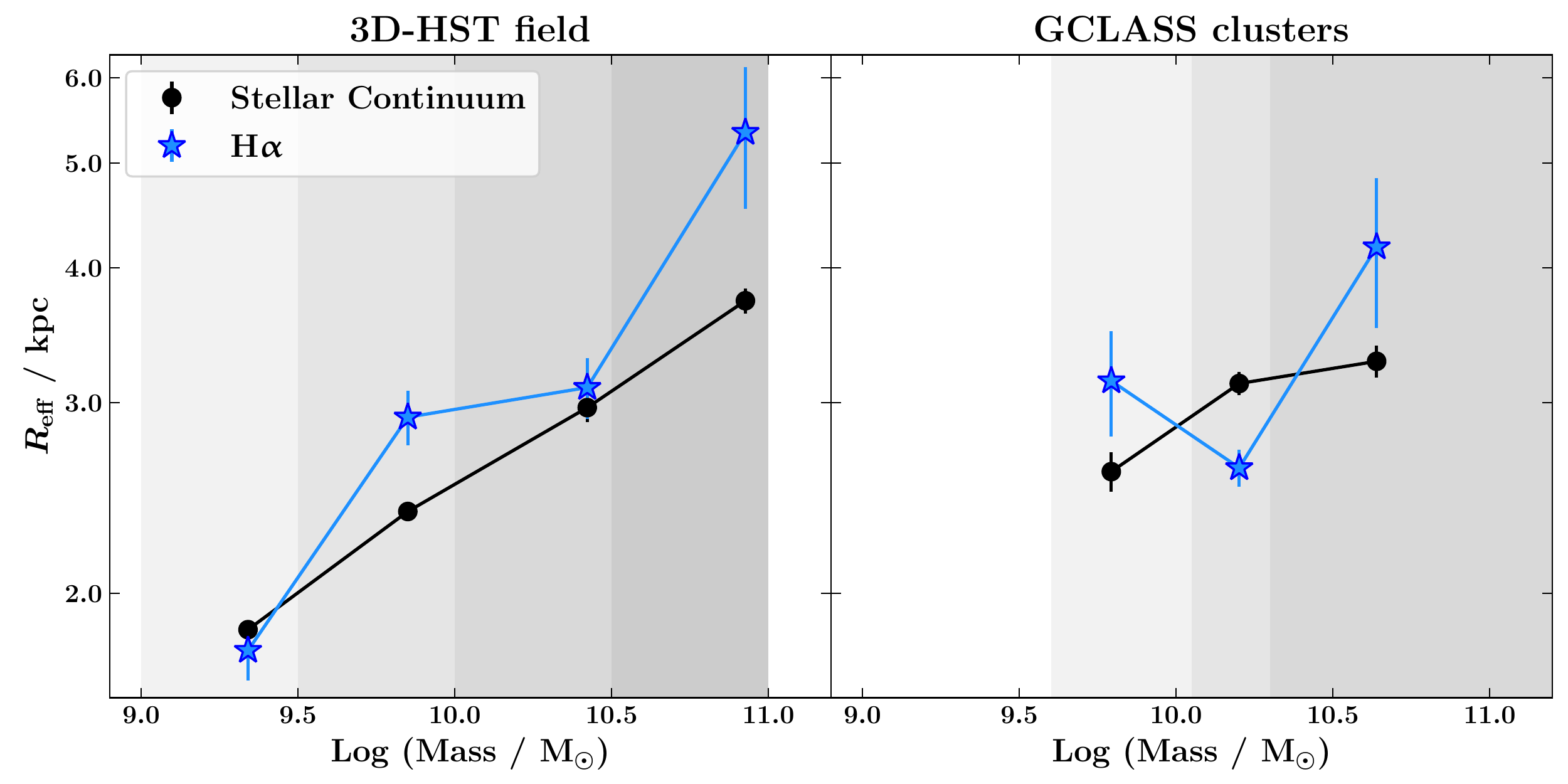}
    \caption{Stellar continuum (filled black circles) and H$\upalpha$ (filled blue stars) stellar mass--size relations at $z\sim1$ for the field (left panel) and cluster (right panel) environments. $R_{\mathrm{eff}}$ is the half-light radius measured along the semi-major axis in kiloparsecs. The grey shaded regions in the background delineate the stellar mass bins for each stack. Both field and cluster environments qualitatively follow the same trend, where on average, H$\upalpha$ is larger than the stellar continuum at fixed stellar mass.}
    \label{fig:ms_relations}
\end{figure*}

\begin{figure}

	\centering\includegraphics[width=\columnwidth]{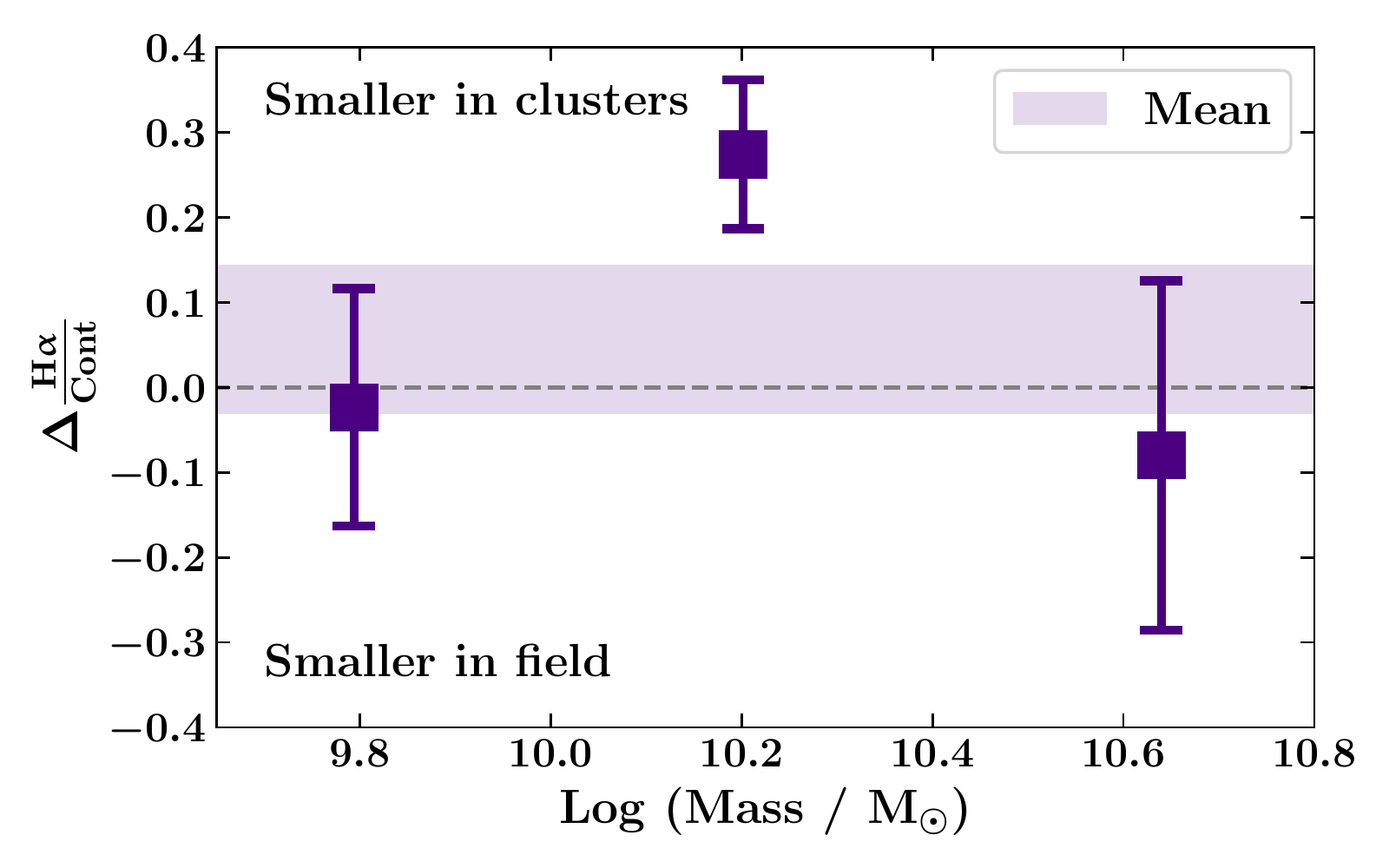}
    \caption{Difference in the ratio between the H$\upalpha$ and stellar continuum spatial extent with environment. \mbox{$\Delta\frac{\mathrm{H}\upalpha}{\mathrm{Cont}} = (\frac{\mathrm{H}\upalpha}{\mathrm{Cont}})_{\mathrm{field}} - (\frac{\mathrm{H}\upalpha}{\mathrm{Cont}})_{\mathrm{cluster}}$}. Values are calculated at the stellar masses of the cluster stacks. The horizontal dashed line delineates no difference with environment. The mean ratio between the H$\upalpha$ and stellar continuum spatial extent is consistent with environment, being smaller in the clusters by $0.06\pm0.09$ over the stellar mass range probed.}
    \label{fig:delta_plot}
\end{figure}

We take the final half-light radii measurements for each stack output by GALFIT in pixels and convert them to kiloparsecs. To do this, we first find the median redshift of the galaxies within each stack. We then calculate the half-light radius, $R_{\mathrm{eff}}$, for each stack in kiloparsecs using the angular diameter distance to this redshift. Similarly, we also find the median stellar mass for the galaxies in each stack. The stellar mass--size relation for the F$140$W and H$\upalpha$ stacks can be seen in the right panel of Figure~\ref{fig:ms_relations}. In the left panel of Figure~\ref{fig:ms_relations}, we show the corresponding GALFIT results from the comparative field environment study using the 3D-HST survey \citep{Nelson2015}. 

Unlike in \cite{Nelson2015} where the errors are calculated from bootstrap resampling the stacks, we calculate our errors by jackknife resampling the stacks. This is due to the expensive nature of the two-GALFIT-run approach (see Section~\ref{size_measurements}) which sometimes requires individual fits to be altered and re-run by hand. Nevertheless, both methods provide a reasonable estimate of the measurement error due to the stacking and size determination process.

Qualitatively, it can be seen that in both environments, the general trend of H$\upalpha$ having a larger half-light radius than the stellar continuum at fixed stellar mass is prevalent (Figure~\ref{fig:ms_relations}). Quantitatively, we show the difference in the ratio of the H$\upalpha$ to stellar continuum half-light radius with environment in Figure~\ref{fig:delta_plot} at the stellar masses of the cluster stacks. The corresponding H$\upalpha$ and stellar continuum measurements in the field are found by linearly interpolating the \cite{Nelson2015} results in log space. Over the stellar mass range probed, we find that on average, the ratio of the H$\upalpha$ to stellar continuum half-light radius is formally smaller in the clusters by $(6\pm9)\%$ (horizontal purple shaded region), but consistent with environment within the $1\sigma$ uncertainties.

\section{Discussion}
\label{discussion}
\subsection{The lack of H\texorpdfstring{$\alpha$}{TEXT} size difference between cluster and field environments}
\label{lack}

In Section~\ref{results}, we measured the half-light radii of $z\sim1$ cluster and field star-forming galaxies in H$\upalpha$ and the stellar continuum, finding that the H$\upalpha$ half-light radius is larger than that of the stellar continuum in both environments (Figure~\ref{fig:ms_relations}) and by the same amount (Figure~\ref{fig:delta_plot}) over the stellar mass range probed. This result suggests that inside-out growth via star formation (traced by H$\upalpha$ emission) is proceeding at the same rate in star-forming galaxies regardless of the environment they reside in. 

Detailed low redshift observations of star-forming cluster galaxies -- particularly in the Virgo cluster -- have provided us with a good understanding of how environmental quenching mechanisms effect the spatial distribution of H$\upalpha$ emission in these galaxies. These works have confirmed that ram-pressure stripping \citep{Gunn1972} for example quenches galaxies from the ``outside-in", leaving the overall stellar distribution intact, but forces star formation -- and therefore H$\upalpha$ emission -- to become progressively confined towards the central regions of galaxies \citep{Hodge1983,Athanassoula1993,Ryder1994,Kenney1999,Koopmann2004,Koopmann2006,Cortes2006,Crowl2006,Koopmann2004,Abramson2011a,Vollmer2012,Gavazzi2013,Ebeling2014,Kenney2015,Abramson2016b,Lee2017,Bellhouse2017,Gullieuszik2017,Sheen2017,Gavazzi2018,Fossati2018,Cramer2019,Boselli2020}. This disk truncation is seen clearly in {\it individual} star-forming cluster galaxies at low redshift, particularly with integral field spectroscopy of ``Jellyfish" galaxies as part of the GASP survey \citep{Poggianti2017,Bellhouse2017,Gullieuszik2017,Fritz2017,Moretti2018a,Jaffe2018,Vulcani2018, Poggianti2019, Poggianti2019a, Vulcani2020} and narrow-band imaging with long-slit spectroscopy in the VESTIGE \citep{Fossati2018,Boselli2020,Boselli2021} survey.

If outside-in quenching via ram-pressure stripping is occurring in $z\sim1$ clusters, we should be able to measure the same H$\upalpha$ disk truncation as is measured in local cluster galaxies from our H$\upalpha$ stacks. We find that the ratio of the H$\upalpha$ to stellar continuum half-light radius in $z\sim1$ star-forming galaxies is smaller in the cluster environment by only $(6\pm9)\%$. Taken at face value, this result is statistically consistent with the coeval field environment, implying outside-in quenching is not occurring in clusters at $z\sim1$, is a sub-dominant environmental quenching mechanism at this epoch, or is occurring rapidly.

\subsection{Evidence for rapid outside-in quenching}
\label{outsidein}

\begin{figure*}

	\centering\includegraphics[width=\textwidth]{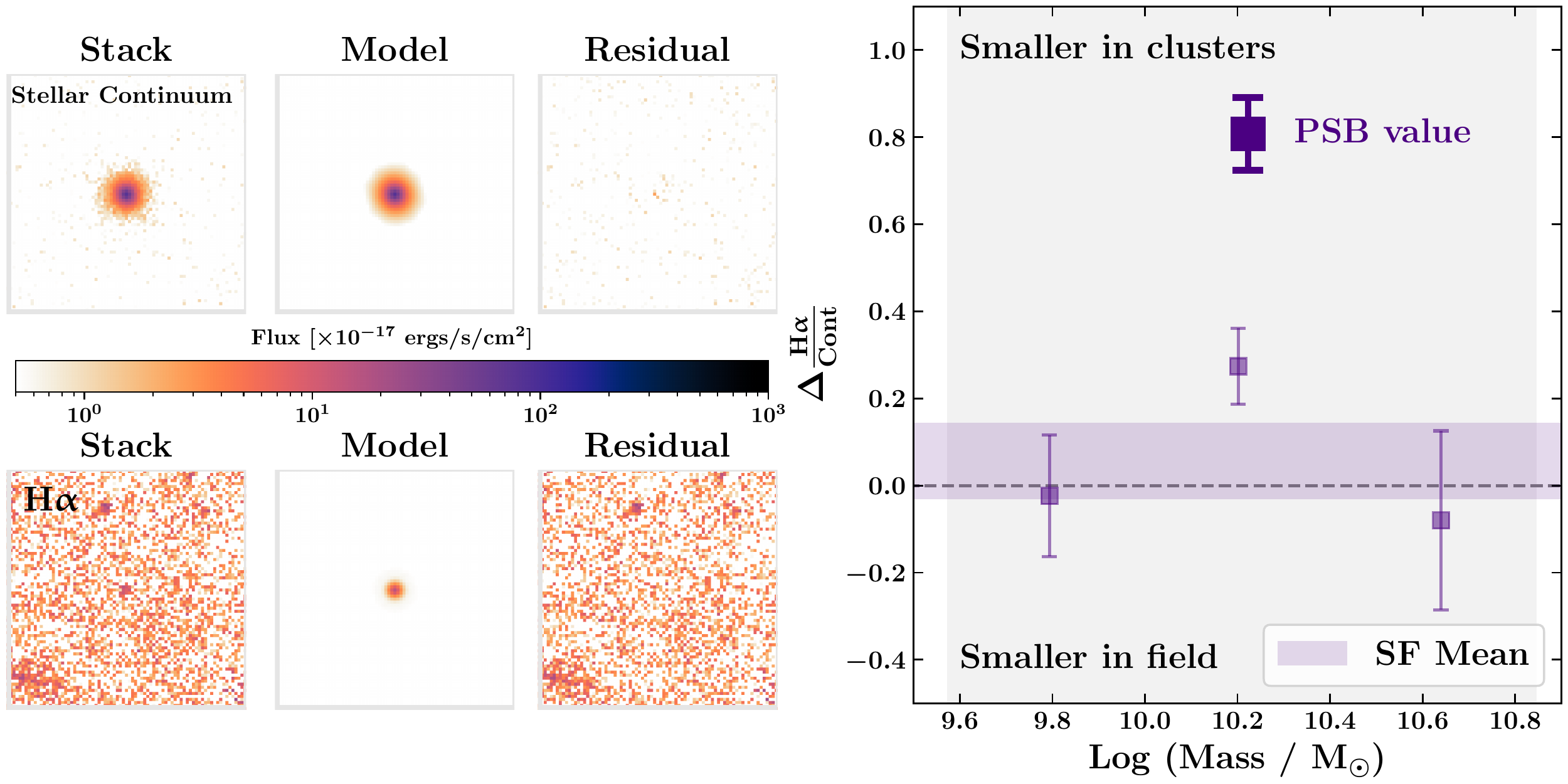}
    \caption{Stellar Continuum versus H$\upalpha$ results for the GCLASS poststarburst (PSB) cluster galaxies. Left panel: F140W Stellar Continuum and G141 H$\upalpha$ stacks with associated GALFIT fits of the 20 PSBs with G141 H$\upalpha$ maps. The colormap is logarithmic, with H$\upalpha$ stacks and fits multiplied by a factor 100 for visibility. Right panel:
    \mbox{$\Delta\frac{\mathrm{H}\upalpha}{\mathrm{Cont}} = (\frac{\mathrm{H}\upalpha}{\mathrm{Cont}})_{\mathrm{field}} - (\frac{\mathrm{H}\upalpha}{\mathrm{Cont}})_{\mathrm{cluster}}$}. The $\Delta\frac{\mathrm{H}\upalpha}{\mathrm{Cont}}$ between the field and cluster star-forming galaxies from Figure~\ref{fig:delta_plot} is shown as the small purple square markers with shaded purple region showing the mean. The $\Delta\frac{\mathrm{H}\upalpha}{\mathrm{Cont}}$ between the field star-forming galaxies and cluster PSB galaxies at the median stellar mass of the PSBs is shown as the large square marker. The ratio of the H$\upalpha$ to the stellar continuum half-light radius for the PSBs is $(81\pm8)\%$ smaller than that for star-forming galaxies with the same stellar mass in the field environment.}
    \label{fig:psb}
\end{figure*}

We know quenching is occurring in the GCLASS clusters and more effectively than in the coeval field, due to the high quenched fractions in the clusters (quenched fraction at Log$(M_{*}/\mathrm{M}_{\odot}) \geqslant 10$ is $50\%$ in the clusters versus $30\%$ in the coeval field) and the overabundance of recently and rapidly quenched (or ``poststarburst", PSB) galaxies in the clusters \citep{Muzzin2012}. 

Over the past decade, a consensus has emerged that environmental quenching likely follows the ``delayed-then-rapid" model proposed by \cite{Wetzel2013}.
In this model, galaxies continue forming stars as they are accreted on to the dark matter halo of another galaxy, but they do not immediately have lower SFRs compared to central galaxies of the same mass. This suggests that quenching does not immediately begin once a galaxy becomes a satellite, but that there is a ``delay" in quenching. During the ``delay" phase, there is a gradual decline in SFR. A ``rapid" catastrophic event then quenches the satellite galaxy, making it difficult to measure any change in the average SFR. \cite{Wetzel2013} proposed this model to reconcile the high quenched fractions in $z=0$ clusters with an environmentally independent SFMS. High redshift studies have also seen a similar independence of the SFMS with environment (e.g., Figure~\ref{fig:SFMS} and \citealt{Muzzin2012, Koyama2013, Old2020,Nantais2020}) and within this framework of environmental quenching have found that the delay time of environmental quenching is shorter than it is at $z=0$ \citep{Muzzin2014a,Balogh2016a,Foltz2018}. \cite{Muzzin2014a} found a delay time a factor $\sim2$ times shorter than that measured at $z=0$ for the GCLASS clusters.

How do we then reconcile the high quenched fractions, overabundance of PSBs \citep{Muzzin2012} and short delay times \citep{Muzzin2014a} in the GCLASS clusters with the indistinguishable H$\upalpha$-to-continuum size ratio with the coeval field (Figure~\ref{fig:delta_plot})? Since cluster galaxies are quenching from the outside-in at low redshift (Section~\ref{lack}), but may be doing so more rapidly at $z\sim1$, it makes sense to examine the recently quenched galaxies in the GCLASS clusters.

Under the hypothesis that delay times for environmental quenching are shorter at $z\sim1$, our stacks (Figure~\ref{fig:fits}) would be dominated by star-forming cluster galaxies yet to undergo environmental quenching, washing out any potential outside-in quenching signal. This is because longer delay times allow for a measurable reduction in the H$\upalpha$ disk size, since a larger fraction of cluster star-forming galaxies can be ``caught-in-the-act" of environmentally quenching. Shorter delay times lead to cluster galaxies dropping out of the star-forming population more quickly, leaving only the star-forming cluster galaxies yet to undergo environmental quenching dominating the stacks. In support of the ``outside-in" environmental quenching scenario, previous studies on the GCLASS clusters which have focused on the spectroscopically confirmed ``recently quenched" or ``poststarburst" (PSB) cluster galaxy population have found evidence for rapid environmental quenching \citep{Muzzin2012,Muzzin2014a} leading to disk truncation in the stellar continuum \citep{Matharu2020}. These studies point towards ram-pressure stripping being the likely quenching mechanism, acting within $0.4_{-0.4}^{0.3}$~Gyr after satellite cluster galaxies make their first passage of $0.5R_{200}$, leaving the stellar component of the galaxy morphologically undisturbed (symmetrical, regular morphologies) but truncating its measured F140W half-light radius.

We perform the same analysis conducted on the star-forming cluster galaxies in this paper on the spectroscopically confirmed recently and rapidly quenched cluster galaxies. 20 of the 23 PSBs studied in \cite{Matharu2020} surprisingly have H$\upalpha$ emission detected in their grism spectra by \texttt{Grizli}. Based on their \texttt{Grizli}-determined H$\upalpha$ fluxes, the mean SFR of the PSBs is $0.7\pm0.1~\mathrm{M_{\odot}}\mathrm{yr}^{-1}$. This SFR is $8\pm1$ times lower than the mean SFR of star-forming cluster galaxies ($6\pm0.1~\mathrm{M_{\odot}}\mathrm{yr}^{-1}$) at a similar stellar mass (middle bin, Figure~\ref{fig:SFMS}). These SFRs are calculated using the mean of the \texttt{Grizli}-determined H$\upalpha$ fluxes from individual detections. Due to the low SNR of the PSB H$\upalpha$ stack, we are unable to fit for both a half-light radius and S\'ersic index simultaneously. We therefore follow the same size determination method as was used for the main analysis in this paper (Section~\ref{size_measurements}), but fix the S\'ersic index to the value determined for the H$\upalpha$ stack of the star-forming cluster galaxies at almost the same stellar mass as the median stellar mass of the PSBs. This is the H$\upalpha$ stack for the second stellar mass bin in the main analysis of the paper. The resulting fits are shown in the left panel of Figure~\ref{fig:psb}.

We do indeed find that the half-light radius of the H$\upalpha$ emission is significantly smaller, by $(70\pm4)\%$, than that of the stellar continuum. We caution the reader, however, that the error bar on the H$\upalpha$ measurement may be underestimated, due to 4 of the 20 jackknife resampled H$\upalpha$ stacks obtaining unreliable size measurements. The ratio of the H$\upalpha$ to stellar continuum half-light radius for the PSBs is $(81\pm8)\%$ smaller than for coeval star-forming field galaxies of the same stellar mass (Right panel, Figure~\ref{fig:psb}). This rather significant difference in the H$\upalpha$ to stellar continuum half-light radius ratio with environment for {\it rapidly} quenched cluster galaxies is consistent with the short delay times measured at high redshift in the delayed-then-rapid quenching model framework \citep{Muzzin2014a,Balogh2016a,Foltz2018}. It further suggests that most of the star-forming cluster galaxies in our main analysis are yet to undergo environmental quenching, and that significant changes in the disk occur just prior to or during the PSB phase. Interestingly, the PSB H$\upalpha$ disks are not only fainter in H$\upalpha$, but also smaller than their star-forming counterparts in the coeval field, implying environmental quenching does operate in an ``outside-in" manner as seen in the local Universe.

We therefore conclude that disk truncation due to ram-pressure stripping is occurring in cluster galaxies at $z\sim1$, and is consistent with operating over shorter delay times than observed in $z\sim0$ clusters. This leads to the effects on the H$\upalpha$ distribution more prominently seen  in the recently and rapidly quenched cluster galaxy population. Deeper, higher SNR H$\upalpha$ emission line maps of the PSBs from grism spectroscopy will be required for a more detailed comparison of the properties of star-forming disks in cluster star-forming and PSB galaxies. Moreover, there may be a small population of star-forming disks that are in a phase prior to the PSB phase and quenching outside-in.  If so, deeper observations providing high SNR H$\upalpha$ emission line maps of {\it individual} galaxies would be key for identifying this population.

\subsection{Literature comparison}
\label{lit_comp}
Our interpretation of the results (Section~\ref{outsidein}) fits well with the growing picture of environmental quenching in the literature, where the delay time of environmental quenching is found to decrease with both increasing redshift and host halo mass (see \citealt{Foltz2018} and references therein). Furthermore, recent results from the GOGREEN survey \citep{Balogh2020a} -- which includes 5 of the 10 GCLASS clusters used in our study -- show a higher quiescent fraction in clusters with respect to the coeval field and an indistinguishable star-forming galaxy stellar mass function (SMF) with environment \citep{VanderBurg2020}. Similar results to those in GOGREEN were also found in more moderate overdensities at similar redshifts as part of the ZFOURGE and NEWFIRM Medium-Band surveys \citep{Papovich2018}. Higher quiescent fractions in clusters and an indistinguishable star-forming galaxy SMF can also be explained by short delay times for environmental quenching and/or high environmental quenching rates building up a significant quiescent fraction in clusters at $z\sim1$. Short delay times would lead to very few star-forming cluster galaxies caught-in-the-act of environmentally quenching (Section~\ref{outsidein}). High environmental quenching rates -- regardless of how long or short the delay time is -- would lead to the build up of a large quiescent fraction in the clusters.

\begin{figure}

	\centering\includegraphics[width=\columnwidth]{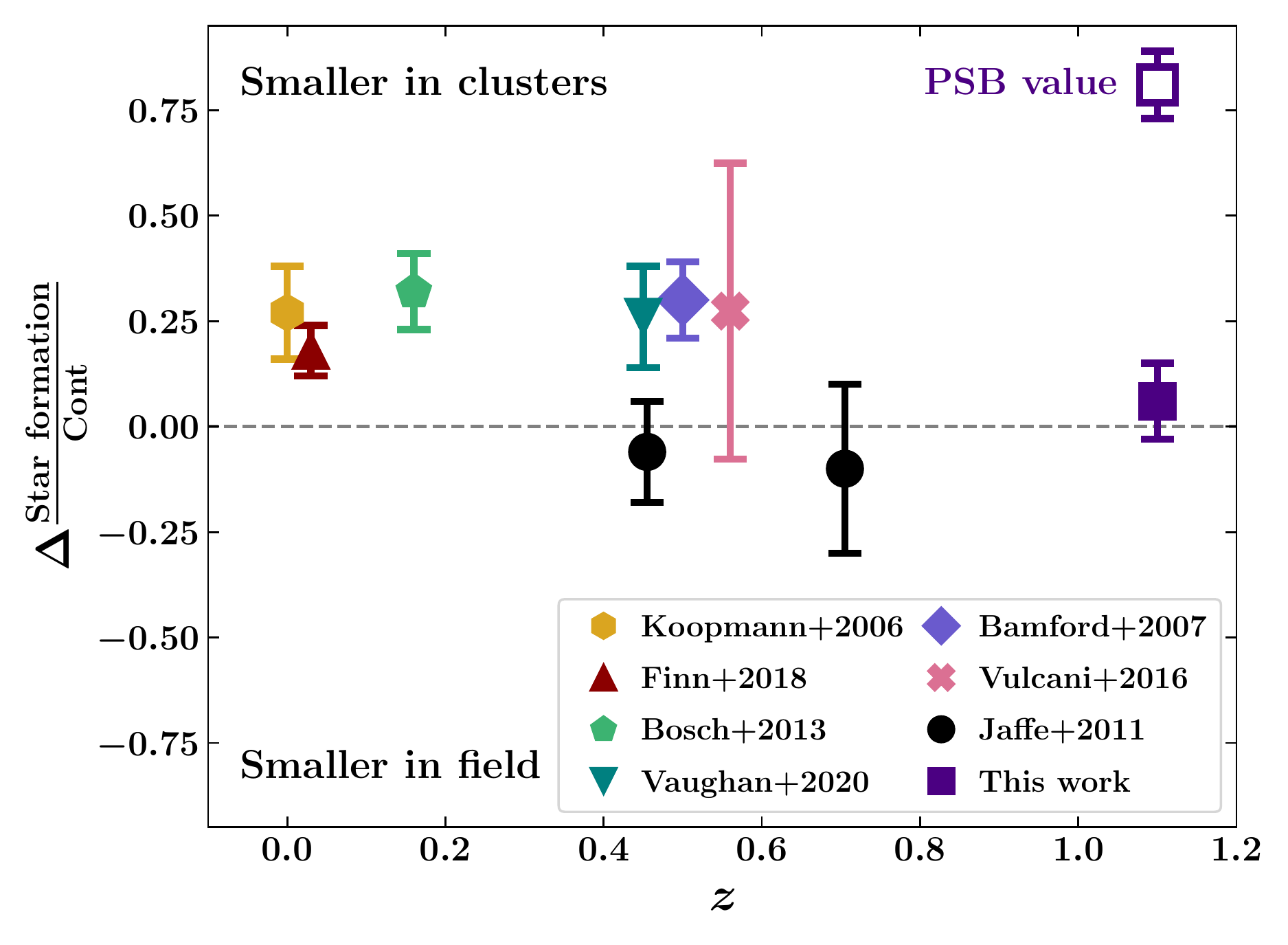}
    \caption{Literature compilation of results from studies which measure the spatial extent of emission from current star formation and from the integrated star formation history as a function of environment. Legend labels are ordered by ascending redshift. The \cite{Vulcani2016} result has been offset from $z\sim0.5$ for clarity. Disk truncation in star-forming cluster galaxies has been measured at $z\lesssim0.5$. We measure significant disk truncation in cluster poststarburst galaxies at $z\sim1$ (open square marker) but not in cluster star-forming galaxies at $z\sim1$ (filled square marker). See Section~\ref{lit_comp} for more details.}
    \label{fig:lit_results}
\end{figure}

Possible evidence for outside-in environmental quenching operating over longer delay times emerges towards lower redshifts in the literature, using a similar technique to the one used in our study. Figure~\ref{fig:lit_results} shows a compilation of results from studies that have measured the spatial extent of current star formation versus the spatial extent of the integrated star formation history as a function of environment using various tracers. The most recent result of disk truncation in star-forming cluster galaxies outside the local Universe was reported from the K-CLASH survey at $0.3<z<0.6$, finding the H$\upalpha$ half-light radius to be smaller than the continuum half-light radius in clusters by $(26\pm12)\%$ \citep{Vaughan2020}. Similar results with somewhat increasing significance have been reported at $z\leqslant0.5$, supporting the possibility that environmental quenching processes that quench cluster galaxies from the outside-in operate over longer delay times at lower redshifts \citep{Koopmann2006,Bamford2007,Jaffe2011,Cortese2012,Bosch2013,Bretherton2013,Vulcani2016,Schaefer2017,Finn2018,Vaughan2020}. 

The levels of disk truncation measured in cluster star-forming and PSB galaxies at $z\sim1$ straddle the range of disk truncation levels one can expect to measure from outside-in environmental quenching. Therefore, if outside-in quenching operates over longer delay times at lower redshift, we would expect to see levels of disk truncation measured that fall somewhere in between the values measured for star-forming and PSB cluster galaxies at $z\sim1$. This indeed seems to be the case when comparing the low redshift literature values to our work (Figure~\ref{fig:lit_results}). However, it is important to note that many of these studies use different tracers and methods to measure the spatial extent of current star formation versus the integrated star formation history. Furthermore, many define environment and select their sample of star-forming galaxies differently over a variety of stellar mass ranges.

Previous results at low redshift ($z\leqslant0.5$) show that the star-forming disks relative to continuum disks in star-forming galaxies are smaller in clusters by $18-32\%$ (Figure~\ref{fig:lit_results}). Our HST data, while requiring stacking, has a 9\% $1\upsigma$ error bar. Therefore, we can rule out star-forming disks being $18-32\%$ smaller than their continuum disks in $z\sim1$ star-forming cluster galaxies at $2-3\upsigma$. Hence if there was no change in how outside-in environmental quenching operates out to $z\sim1$, our data have the sensitivity to confirm such a hypothesis. Instead, we see a factor of 3 evolution in disk truncation measurements between $z\sim0$ and $z\sim1$ (Figure~\ref{fig:lit_results}, and see \citealt{Noble2019} for similar levels of evolution measured in CO molecular gas to stellar continuum half-light radius ratios at $z\sim1.6$), implying that the outside-in environmental quenching process evolves between $0\leqslant z \leqslant1$. More specifically, the disk truncation measurements for star-forming cluster galaxies show that outside-in quenching is either a sub-dominant environmental quenching mechanism at $z\sim1$ or is operating over shorter delay times. The PSB result at $z\sim1$ confirms outside-in quenching is operating in $z\sim1$ clusters. Combining the PSB result with the known overabundance of PSBs \citep{Muzzin2012} and shorter delay times \citep{Muzzin2014a} in the GCLASS clusters suggests shorter delay times for outside-in environmental quenching and/or higher outside-in environmental quenching rates at $z\sim1$ compared to $z\sim0$.

\section{Summary}
\label{summary}

By carefully conducting HST WFC3 F140W and G141 grism observations in the same way as was done for 3D-HST (Section~\ref{HST_data}), we have made the first attempt in measuring an outside-in environmental quenching signal in the largest sample of spectroscopically-confirmed star-forming cluster galaxies at $z\sim1$ (Section~\ref{results}).

To further reduce systematics and improve measurements, we processed and stacked our data using the 3D-HST methodology, but used improved methods where our more sophisticated software and methods provided them (Section~\ref{methodology}).

For the benefit and use of the scientific community, we publicly release all processed grism data (Section~\ref{public_data_release} and Appendix~\ref{appendix_release}) for the 10 GCLASS clusters this study is based on at the website \url{https://www.gclasshst.com}.

Our main conclusions are as follows:
\begin{enumerate}

\item Qualitatively, both field and cluster environments at $z\sim1$ follow the same trend in stellar continuum and H$\upalpha$ half-light radius, where H$\upalpha$ is on average larger than the stellar continuum at fixed stellar mass (Figure~\ref{fig:ms_relations}).

\item Quantitatively, the ratio of the H$\upalpha$ to stellar continuum half-light radius for star-forming galaxies is smaller in the cluster environment by $(6\pm9)\%$ but statistically consistent with being the same in all environments at $z\sim1$ (Figure~\ref{fig:delta_plot}).

\item Given the high quenched fractions in the clusters compared to the coeval field, the overabundance of PSBs in the clusters and the similarity in star-forming SMFs, our results are consistent with environmental quenching being a rapid process at $z\sim1$, explaining the similarity we see in the H$\upalpha$ and stellar continuum mass--size trends of star-forming field and cluster galaxies (Section~\ref{outsidein}).

\item By examining a population of cluster galaxies spectroscopically identified as recently and rapidly quenched (also known as ``poststarbursts", PSBs), we find that 87\% of these galaxies have H$\upalpha$ detections, implying SFRs of $0.7\pm0.1~\mathrm{M_{\odot}}\mathrm{yr}^{-1}$ (Section~\ref{outsidein}). This SFR is $8\pm1$ times lower than the SFR of star-forming cluster galaxies with similar stellar masses.

\item As well as having very low SFRs, PSB cluster galaxies have H$\upalpha$ half-light radii $(70\pm4)\%$ smaller than the half-light radii of their stellar continuum. When compared to star-forming field galaxies at fixed stellar mass, PSB cluster galaxies have a H$\upalpha$ to stellar continuum half-light radius ratio that is $(81\pm8)\%$ smaller (Right panel, Figure~\ref{fig:psb}). This result suggests that the rapid environmental quenching process responsible for creating these galaxies propagated in an outside-in manner, severely truncating their H$\upalpha$ disks.

\item The sensitivity of our data allow us to confirm a factor 3 evolution in disk truncation levels from outside-in quenching between $0\leqslant z \leqslant1$ (Section~\ref{lit_comp} and Figure~\ref{fig:lit_results}), consistent with the sub-dominance of outside-in environmental quenching at high redshift, the shortening of delay times for outside-in environmental quenching at high redshift and/or an increase in outside-in environmental quenching rates at high redshift.

\end{enumerate}

Our work shows that the environmental quenching process evolves with cosmic time, calling for the need to rethink existing simplistic models of time-independent environmental quenching.

Moving forward, deeper, higher SNR H$\upalpha$ emission line maps of star-forming and PSB cluster galaxies will be required to explore the disk structures in more detail. This will allow us to measure the scatter in their physical properties and move away from stacking analyses which are notorious for diluting important physical signals. The future is promising in this endeavour, thanks to the many slitless and even integral-field spectroscopy capabilities on-board the {\it James Webb Space Telescope} (JWST) and the {\it Nancy Grace Roman Space Telescope} (NGRST). 

\acknowledgments

\noindent JM thanks Raymond Simons, Vicente Estrada-Carpenter and Ivelina Momcheva for providing crucial details regarding the inner workings of \texttt{Grizli} and its data products. JM is also grateful for the support from the George P. and Cynthia Woods Mitchell Institute for Fundamental Physics and Astronomy at Texas A\&M University. This work is supported by the National Science Foundation through grant AST-1517863, AST-1517815, NASA HST grants AR-14310 and GO-15294, and by grant numbers 80NSSC17K0019 and 80NSSC19K0592 issued through the NASA Astrophysics Data Analysis Program (ADAP). Support for program number GO-15294 was provided by NASA through a grant from the Space Telescope Science Institute, which is operated by the Association of Universities for Research in Astronomy, Incorporated, under NASA contract NAS5-26555. GR also acknowledges the support of an ESO visiting science fellowship. The Cosmic Dawn Center of Excellence is funded by the Danish National Research Foundation under grant No. 140. RD gratefully acknowledges support from the Chilean Centro
de Excelencia en Astrof\'isica y Tecnolog\'ias Afines (CATA) BASAL grant
AFB-170002.

\software{This research made use of \textsc{Astropy}, a community-developed core Python package for Astronomy \citep{TheAstropyCollaboration2018}. The python packages \textsc{Matplotlib} \citep{Hunter2007}, \textsc{Numpy}, and \textsc{Scipy} were also extensively used. Parts of the results in this work make use of the colormaps in the \textsc{CMasher} \citep{VanderVelden2020} package.}
\appendix
\label{appendix}

\section{Public Data Release details}
\label{appendix_release}
In this Appendix we provide further details on the data products in the public data release summarized in Section~\ref{public_data_release}. Full details can be found at the data release website, \mbox{\url{https://www.gclasshst.com}}.

\subsection{F140W direct image thumbnails}
\label{direct_image_thumbnails}
The F140W direct image thumbnails (see panel A of Figure~\ref{fig:data_release}) have the same WCS and drizzle parameters as the emission line maps and can therefore be used to perform direct comparisons with the emission line maps (see second sub-panel of panel D, Figure~\ref{fig:data_release}). Note, however, that the direct image thumbnails are in units of electrons s$^{-1}$ and the emission line maps are in units of $\times10^{-17}$~ergs~s$^{-1}$~cm$^{-2}$. 

Within the same fits file as the direct image thumbnails, the user will find a segmentation map thumbnail (see panel A of Figure~\ref{fig:data_release}), an inverse variance (or ``weight") map thumbnail and a WFC3 F140W point-spread function (PSF) thumbnail for the source. Each thumbnail is $80\times80$ pixels, with a pixel scale of 0.1$^{\prime\prime}$.

\subsection{Two-dimensional G141 grism spectra}
\label{2dspec}
There are two data products for the 2D G141 grism spectra of a source. The first of which is a fits file (with file-ending \texttt{beams.fits}) containing each independent G141 grism + F140W imaging observation and their associated calibration images (see website for more details). There are often several observations for each individual position angle (PA) of the G141 grism. Note that the GCLASS G141 observations were taken at one PA and do {\it not} have their contamination and continuum removed in this file.

The second data product provides the stacked 2D G141 grism spectrum for each individual PA and the PA-combined stack (see panel B in Figure~\ref{fig:data_release}). In the case of GCLASS, all G141 observations were taken at a single PA. Therefore the PA-combined stack (bottom row of panel B, Figure~\ref{fig:data_release}) is the same as the stacked individual PA spectrum (top row of panel B, Figure~\ref{fig:data_release}). More details regarding the associated calibration images can be found on the data release website.

\subsection{One-dimensional G141 grism spectra}
\label{1dspec}
The collapsed 1D G141 grism spectra, along with their associated best-fit Spectral Energy Distribution (SED) and calibration files are also available in this data release. An example 1D G141 grism spectrum for a galaxy in our final cluster sample and its best-fit model can be seen in panel C of Figure~\ref{fig:data_release}.

\subsection{Emission line maps}
\label{emaps}
The final data product contains the results of the redshift fit (first plot in panel C, Figure~\ref{fig:data_release}) and the emission line maps for each source (panel D, Figure~\ref{fig:data_release}). Note that the emission line maps already have the continuum and contamination removed, despite the calibration images being available in the same fits file. Each thumbnail is $80\times80$ pixels, with a pixel scale of 0.1$^{\prime\prime}$. See Section~\ref{direct_image_thumbnails} for details regarding the direct image thumbnails in this file, a zoomed-in region of which is shown in the first sub-panel of panel D, Figure~\ref{fig:data_release}.


\bibliography{library_grizli1}{}
\bibliographystyle{aasjournal}



\end{document}